\documentclass[aps,prd,nofootinbib,showpacs]{revtex4}
\usepackage{amssymb}
\usepackage{amsmath}
\usepackage{amsfonts}
\usepackage{epsfig}
\usepackage{appendix}
\def\uu{\langle \bar u u \rangle}
\def\dd{\langle \bar d d \rangle}

\newcommand{\seq}{\begin{subequations}}
\newcommand{\sen}{\end{subequations}}
\newcommand{\eq}{\begin{eqnarray}}
\newcommand{\en}{\end{eqnarray}}

\begin{document}

\title{  \bf Semileptonic $\Lambda_{b,c}$ to Nucleon  Transitions  in Full QCD at Light Cone }
\author{ K. Azizi$^{1,\dag}$,  M. Bayar$^{2,\ddag}$,   Y. Sarac$^{3,*}$, H. Sundu $^{4,\ddag}$\\
 $^\dag$ Physics Division,  Faculty of Arts and Sciences, Do\u gu\c s University,
 Ac{\i}badem-Kad{\i}k\"oy, \\ 34722 Istanbul, Turkey\\
$^\ddag$ Department of Physics, Kocaeli University, 41380 Izmit, Turkey\\
$^*$ Electrical and Electronics Engineering Department, Atilim University, 06836 Ankara, Turkey\\
$^1${e-mail:kazizi@dogus.edu.tr} \\
$^2$melahat.bayar@kocaeli.edu.tr\\
$^3$ysoymak@atilim.edu.tr\\
$^4$hayriye.sundu@kocaeli.edu.tr }

\begin{abstract}
The tree level semileptonic $\Lambda_{b}\rightarrow pl\nu$ and
$\Lambda_{c}\rightarrow nl\nu$ transitions  are investigated using the
light cone QCD sum rules approach in full theory. The spin--1/2, $\Lambda_{Q}$ baryon with  $Q=b$ or $c$, is considered by the
most general form of its interpolating current. The time ordering
product of the initial and transition currents is expanded in terms
of the nucleon distribution amplitudes with different twists.
Considering two sets of independent input parameters entering to the
nucleon wave functions, namely, QCD sum rules and Lattice QCD parameters, the
related form factors and their heavy quark effective theory limits
are calculated and compared with the existing predictions of other
approaches. It is shown that our results satisfy the heavy quark
symmetry relations for lattice input parameters and b case exactly
and the maximum violation is for charm case and QCD sum rules input
parameters. The obtained form factors are used to compute the
transition rates both in full theory and heavy quark effective theory. A comparison of the
results on decay rate of  $\Lambda_{b}\rightarrow pl\nu$ with those  predicted by other
phenomenological methods or the same method in heavy quark effective theory  with different
interpolating current and distribution amplitudes  of the $\Lambda_{b}$  is also
presented.
\end{abstract}


\maketitle

\section{Introduction}
Motivated by  the recent experimental progresses on the spectroscopy
of the heavy baryons containing  heavy b or c quark
\cite{Mattson,Ocherashvili,Acosta,Chistov,Aubert1,Abazov1,Aaltonen1,Solovieva},
theoretical studies on these baryons gain pace. Because of the heavy
quark, these states are expected to be narrow, experimentally, hence
 their isolation and detection are easy comparing with light systems.
 Theoretically, investigation of the  semileptonic decays of the heavy
 baryons, whose experimental testing  may be in the future
program of the large hadron collider (LHC), have attracted
interests beside their mass and electromagnetic properties.  For
instance, the semileptonic $\Lambda_b\rightarrow \Lambda_c$ and
$\Lambda_c\rightarrow \Lambda$ decays have been investigated in
three points QCD sum rules and heavy quark effective theory (HQET)
in \cite{yeni1}. The $\Lambda_{b}\rightarrow pl\bar\nu$ transition
has also been studied in the same frameworks in \cite{yeni2} and
using SU(3) symmetry and HQET in \cite{Datta}. Constituent quark
model have also been used to study the $\Lambda_{c}\rightarrow
nl\bar\nu$ and $\Lambda_{b}\rightarrow pl\bar\nu$ form factors
\cite{pervin} and semileptonic decays of some heavy  baryons
containing single heavy quark in different quark models
\cite{D.Ebert,albertus,pervin} are some other works in this respect.

In our recent work \cite{kmayprd}, we analyzed the semileptonic
decay of $\Sigma_b$, which has different interpolating current and
structure than $\Lambda_{Q}$, to proton in light cone QCD sum rules.
In present study,   we calculate the form factors related to the
semileptonic decays of the $\Lambda_b\rightarrow pl\nu$ and
$\Lambda_c\rightarrow nl\nu$ also in light cone QCD in full theory
and HQET limit. In full theory, these transitions are governed by
six form factors, but heavy quark effective theory limit reduces
them to two. The vacuum to nucleon matrix element of the time
ordering product is expanded in terms of nucleon distribution
amplitudes (DA's) near light cone, $x^2\simeq0$. The nucleon wave
functions contain eight independent parameters, which we consider
two sets, namely, calculated using the QCD sum rules \cite{Lenz} and
lattice QCD \cite{Gockeler1,Gockeler2,QCDSF} approaches. In the
calculations, the most general current  of $\Lambda_{Q}$
generalizing the Ioffe current is used. The obtained form factors
are used to compute the corresponding  transition rates both their
numerical values and in terms of the Cabibbo-Kobayashi-Maskawa (CKM)
matrix elements. Studying such type of transition provides a better
understanding of the internal structure of $\Lambda_Q$, information
about the DA's and input parameters as well as determination of the
CKM matrix elements. Note that, using different interpolating field,
the semileptonic decay of bottom case, $\Lambda_b\rightarrow pl\nu$,
has already been investigated in references \cite{yeni4,yeni5} in
the same framework but HQET limit. In \cite{yeni4}, the nucleon
distribution amplitudes are used only with QCD sum rules input
parameters, while the distribution amplitudes of $\Lambda_b$ have
been  utilized to calculate the form factors in
\cite{yeni5}.

The layout  of the paper is as follows: in section II, the details
of the calculation of the form factors in light cone QCD sum rules
method are presented where the nucleon distribution amplitudes and
the most general form of the interpolating currents for the
$\Lambda_Q$ baryon are used. The heavy quark limit of the form
factors and the relations between the form factors in this limit is
also discussed in this section. Section III comprises numerical
analysis of the form factors and our predictions for the decay rate
obtained in two different ways: first, using the DA's obtained from
QCD sum rules and second, the DA's calculated in lattice QCD. A
comparison of our results on form factors and transition rates with
the existing predictions of other  approaches is also presented in
this section.

\section{ Theoretical Framework}
In this section, following \cite{kmayprd},  we calculate  the form
factors of the $\Lambda_{b} \rightarrow p$ and $\Lambda_{c}
\rightarrow n$ transitions in the framework of the light cone QCD
sum rules and full theory. At quark level, these decays are governed
by the tree level $Q\rightarrow q$ transition, where $Q$ represents
$b~(c)$ quark and $q$ stands for $u~(d)$ quark for
$\Lambda_{b}(\Lambda_{c})$. The effective Hamiltonian responsible
for these transitions at the quark level has the form
\begin{eqnarray}
{\cal H}_{eff} = \frac{G_F}{\sqrt2} V_{qQ} ~\bar q \gamma_\mu
(1-\gamma_5) Q \bar l \gamma^\mu (1-\gamma_5) \nu.
\end{eqnarray}
 To calculate the amplitude, we need to sandwich the above equation between the initial and final states and compute the  matrix
element $\langle N \vert \bar q \gamma_\mu (1-\gamma_5) Q \vert
\Lambda_{Q} \rangle $, which is needed to study the $\Lambda_{Q}
\rightarrow N l \nu$ decay.  The starting point is to consider the
following correlation function
\begin{equation}\label{T}
\Pi_{\mu}(p,q)=i\int d^{4}xe^{iqx}\langle N(p)\mid
T\{J^{tr}_{\mu}(x)\bar J^{\Lambda_{Q}}(0) \}\mid 0\rangle,
\end{equation}
where, $J^{\Lambda_{Q}}$ is interpolating currents of
$\Lambda_{b(c)}$ baryon,  $J^{tr}_{\mu}=\bar
q\gamma_{\mu}(1-\gamma_5)Q$ is transition current and $\langle
N(p)\mid$ represents the nucleon sate, where $p$ denotes the proton
(neutron) momentum and $q=(p+q)-p$ is the transferred momentum.

One further step of the calculation is the saturation of the
correlation function by a tower of hadronic states having the same
quantum numbers as the interpolating currents. The obtained result
from this procedure is called the phenomenological or physical side
of the correlation function.  From the general philosophy of the QCD
sum rules approach, this correlator is also calculated using the
operator product expansion (OPE) in deep Euclidean region. This part
is called the theoretical or QCD  side. Match of these two different
representations of the same correlation function  gives sum rules
for form factors.  To suppress the contribution of the higher states
and continuum, the Borel transformation is applied to both sides of
the sum rules for physical quantities.

Let first calculate the phenomenological part. After the insertion
of the complete set of the initial hadronic state and performing the
integral over x, we obtain the physical side as:
\begin{equation}\label{phys1}
\Pi_{\mu}(p,q)=\sum_{s}\frac{\langle N(p)\mid J^{tr}_{\mu}(x)\mid
\Lambda_{Q}(p+q,s)\rangle\langle \Lambda_{Q}(p+q,s)\mid \bar
J^{\Lambda_{Q}}(0)\mid 0\rangle}{m_{\Lambda_{Q}}^{2}-(p+q)^{2}}+...,
\end{equation}
where, the ... represents the contribution of the higher states and
continuum. The matrix element $\langle\Lambda_{Q}(p+q,s)\mid \bar
J^{\Lambda_{Q}}(0)\mid 0\rangle$ in (\ref{phys1}) is given by:
\begin{equation}\label{matrixel2}
\langle\Lambda_{Q}(p+q,s)\mid \bar J^{\Lambda_{Q}}(0)\mid
0\rangle=\lambda_{\Lambda_{Q}} \bar u_{\Lambda_{Q}}(p+q,s),
\end{equation}
where $\lambda_{\Lambda_{Q}}$ is residue of $\Lambda_{Q}$ baryon.
The transition  matrix element, $\langle N(p)\mid J_{\mu}^{tr}\mid
\Lambda_{Q}(p+q,s)\rangle$ can be written as

\begin{eqnarray}\label{matrixel1}
\langle N(p)\mid J_{\mu}^{tr}(x)\mid \Lambda_{Q}(p+q)\rangle&=&\bar
N(p)\left[\gamma_{\mu}f_{1}(Q^{2})+{i}\sigma_{\mu\nu}q^{\nu}f_{2}(Q^{2})+
q^{\mu}f_{3}(Q^{2})+\gamma_{\mu}\gamma_5
g_{1}(Q^{2})+{i}\sigma_{\mu\nu}\gamma_5q^{\nu}g_{2}(Q^{2})\right.\nonumber\\
&+& \left. q^{\mu}\gamma_5 g_{3}(Q^{2})
\vphantom{\int_0^{x_2}}\right] u_{\Lambda_{Q}}(p+q),\nonumber\\
\end{eqnarray}
where $Q^{2}=-q^{2}$. The $f_{i}$ and $g_{i}$  are  transition form
factors in full theory and $N(p)$ and $u_{\Lambda_{Q}}(p+q)$ are the
spinors of nucleon and $\Lambda_{Q}$, respectively.  Using Eqs.
(\ref{phys1}), (\ref{matrixel2}) and (\ref{matrixel1}) and summing
over spins of the $\Lambda_{Q}$ baryon, i.e.,
\begin{equation}\label{spinor}
\sum_{s}u_{\Lambda_{Q}}(p+q,s)\overline{u}_{\Lambda_{Q}}(p+q,s)=\not\!p+\not\!q+m_{\Lambda_{Q}},
\end{equation}
 we attain the following expression
\begin{eqnarray}\label{phys2}
\Pi_{\mu}(p,q)&=&
\frac{\lambda_{\Lambda_{Q}}}{m_{\Lambda_{Q}}^{2}-(p+q)^{2}}\bar
N(p)\left[\gamma_{\mu}f_{1}(Q^{2})+{i}\sigma_{\mu\nu}q^{\nu}f_{2}(Q^{2}+
q^{\mu}f_{3}(Q^{2})+\gamma_{\mu}\gamma_5
g_{1}(Q^{2})+{i}\sigma_{\mu\nu}\gamma_5q^{\nu}g_{2}(Q^{2})\right.\nonumber\\
&+& \left. q^{\mu}\gamma_5 g_{3}(Q^{2})
\vphantom{\int_0^{x_2}}\right] (\not\!p+\not\!q+m_{\Lambda_{Q}}) +
\cdots
\end{eqnarray}
 Using
\begin{eqnarray}\label{sigma}
\bar{N}\sigma_{\mu\nu}q^{\nu}u_{\Lambda_{Q}}&=i&
\bar{N}[(m_N+m_{\Lambda_{Q}})\gamma_{\mu}-(2p+q)_\mu]u_{\Lambda_{Q}},
\end{eqnarray}
in Eq. (\ref{phys2}), the following final expression  for the
physical side of the correlation function is obtained:
\begin{eqnarray}\label{sigmaafter}
\Pi_{\lambda}(p,q)&=&
\frac{\lambda_{\Lambda_{Q}}}{m_{\Lambda_{Q}}^{2}-(p+q)^{2}}\bar
N(p)\left[\vphantom{\int_0^{x_2}}2f_{1}(Q^{2})p_\mu+\left\{\vphantom{\int_0^{x_2}}-f_1(Q^2)(m_N-m_{\Lambda_{Q}})
+f_2(Q^2)(m_N^2-m_{\Lambda_{Q}}^2)\right\}\gamma_\mu\right.\nonumber \\
&&+\left\{\vphantom{\int_0^{x_2}}f_1(Q^2)-f_2(Q^2)(m_N+m_{\Lambda_{Q}})\right\}\gamma_\mu\not\!q+
2f_{2}(Q^{2})p_\mu\not\!q
+\left\{\vphantom{\int_0^{x_2}}f_2(Q^2)+f_3(Q^2)\right\}(m_N+m_{\Lambda_{Q}})q_\mu\nonumber\\&+&\left\{\vphantom{\int_0^{x_2}}f_2(Q^2)
+f_3(Q^2)\right\}q_\mu\not\!q- 2g_1(Q^2)p_{\mu}\gamma_5
+\left\{\vphantom{\int_0^{x_2}}g_1(Q^2)(m_N+m_{\Lambda_{Q}})\right.-\left.g_2(Q^2)(m_N^2-m^2_{\Lambda_{Q}})\vphantom{\int_0^{x_2}}\right\}
\gamma_\mu\gamma_5-\nonumber\\&&
\left\{\vphantom{\int_0^{x_2}}g_{1}(Q^{2})-g_2(Q^2)(m_N-m_{\Lambda_{Q}})\right\}\gamma_\mu\not\!q\gamma_5-2g_2(Q^2)p_\mu\not\!q\gamma_5
-\left\{\vphantom{\int_0^{x_2}}g_2(Q^2)+g_3(Q^2)\right\}(m_N-m_{\Lambda_{Q}})q_\mu\gamma_5\nonumber\\&&
-\left\{g_2(Q^2)+g_3(Q^2)\vphantom{\int_0^{x_2}}\right\}q_\mu\not\!q\gamma_5\left.\vphantom{\int_0^{x_2}}\right]+
\cdots
\end{eqnarray}

In order to calculate the form factors $f_{1}$, $f_{2}$, $f_{3}$,
$g_{1}$, $g_{2}$ and  $g_{3}$, we will choose the independent
structures $p_{\mu}$, $p_{\mu}\!\!\not\!q$, $q_{\mu}\!\!\not\!q$,
$p_{\mu}\gamma_5$, $p_{\mu}\!\!\not\!q\gamma_5$, and
$q_{\mu}\!\!\not\!q\gamma_5$ from
 Eq. (\ref{phys2}), respectively.

For the theoretical side, to evaluate the correlation function in
deep Euclidean region where $(p+q)^2\ll0$, the explicit expression
of the interpolating field of the $\Lambda_{Q}$ baryon is needed.
Considering the quantum numbers, the most general form of
interpolating current which can create the $\Lambda_Q$ from the
vacuum is given as

\begin{eqnarray}\label{cur.N}
J^{\Lambda_{Q}}(x)&=&\frac{1}{\sqrt{6}}\epsilon_{abc}\left\{\vphantom{\int_0^{x_2}}2(q_{1}^{aT}Cq_{2}^{b})\gamma_{5}Q^{c}
+\beta(q_{1}^{aT}C\gamma_{5}q_{2}^{b})Q^{c}+(q_{1}^{aT}CQ^{b})\gamma_{5}q_{2}^{c}\right.\nonumber\\
&+&\left.\vphantom{\int_0^{x_2}}
\beta(q_{1}^{aT}C\gamma_{5}Q^{b})q_{2}^{c}+(Q_{}^{aT}Cq_{2}^{b})\gamma_{5}q_{1}^{c}+\beta(Q_{}^{aT}C\gamma_{5}q_{2}^{b})q_{1}^{c}\right\},\nonumber\\
\end{eqnarray}
where $q_{1}$ and $q_{2}$ are the $u$ and $d$ quarks, respectively,
$a,~b,~c$ are the color indices and $C$ is the charge conjugation
operator and $\beta$ is an arbitrary parameter with $\beta=-1$
corresponding to the Ioffe current. Using the transition current,
$J^{tr}_{\mu}=\bar q\gamma_{\mu}(1-\gamma_5)Q$ and $J^{\Lambda_{Q}}
$and contracting out all quark pairs by the help of the Wick's
theorem, we achieve
\begin{eqnarray}\label{mut.m}
\Pi_\mu &=& \frac{-i}{\sqrt{6}} \epsilon^{abc}\int d^4x e^{iqx}
\Bigg\{\Big[2 ( C )_{\eta\phi} (\gamma_5)_{\rho\beta}+( C
)_{\eta\beta} (I)_{\rho\delta}(\gamma_5)_{\delta\phi}+( C
)_{\beta\phi} (\gamma_5)_{\eta\rho}\Big] +\beta\Bigg[2 (C \gamma_5
)_{\eta\phi}(I)_{\rho\beta}
 \nonumber \\
&+& (C \gamma_5 )_{\eta\beta}(I)_{\rho\phi}+(C \gamma_5
)_{\beta\phi}(I)_{\eta\rho} \Bigg]\Bigg\} \Big[
(1+\gamma_5)\gamma_{\mu} \Big]_{\sigma\theta}S_Q(-x)_{\beta\sigma}
\langle  N (p) | \bar u_\eta^a(0)
\bar u_\theta^b(x)  \bar d_\phi^c(0) | 0\rangle ,\nonumber\\
\end{eqnarray}
where, $ S_Q(x)$ is the heavy quark propagator which is given by
\cite{Balitsky}:

\begin{eqnarray}\label{heavylightguy}
 S_Q (x)& =&  S_Q^{free} (x) - i g_s \int \frac{d^4 k}{(2\pi)^4}
e^{-ikx} \int_0^1 dv \Bigg[\frac{\not\!k + m_Q}{( m_Q^2-k^2)^2}
G^{\mu\nu}(vx) \sigma_{\mu\nu} + \frac{1}{m_Q^2-k^2} v x_\mu
G^{\mu\nu} \gamma_\nu \Bigg].
 \end{eqnarray}
where
\begin{eqnarray}\label{freeprop}
S^{free}_{Q}
&=&\frac{m_{Q}^{2}}{4\pi^{2}}\frac{K_{1}(m_{Q}\sqrt{-x^2})}{\sqrt{-x^2}}-i
\frac{m_{Q}^{2}\not\!x}{4\pi^{2}x^2}K_{2}(m_{Q}\sqrt{-x^2}),\nonumber\\
\end{eqnarray}
and  $K_i$ are the Bessel functions. Here, we neglect the terms
proportional to the gluon field strength tensor since they can give
contribution to four and five particle distribution functions and
expected to be small \cite{17,18,Braun1b}.

The matrix element $\langle  N (p)\mid \epsilon^{abc}\bar
u_{\eta}^{a}(0)\bar u_{\theta}^{b}(x)\bar d_{\phi}^{c}(0)\mid
0\rangle$  appearing in Eq. (\ref{mut.m}), which is the nucleon wave
function, is represented as \cite{Lenz,17,18,Braun1b,8}:

\begin{eqnarray}\label{wave func}
&&4\langle0|\epsilon^{abc}u_\alpha^a(a_1 x)u_\beta^b(a_2
x)d_\gamma^c(a_3 x)|N(p)\rangle\nonumber\\
&=&\mathcal{S}_1m_{N}C_{\alpha\beta}(\gamma_5N)_{\gamma}+
\mathcal{S}_2m_{N}^2C_{\alpha\beta}(\rlap/x\gamma_5N)_{\gamma}\nonumber\\
&+& \mathcal{P}_1m_{N}(\gamma_5C)_{\alpha\beta}N_{\gamma}+
\mathcal{P}_2m_{N}^2(\gamma_5C)_{\alpha\beta}(\rlap/xN)_{\gamma}+
(\mathcal{V}_1+\frac{x^2m_{N}^2}{4}\mathcal{V}_1^M)(\rlap/pC)_{\alpha\beta}(\gamma_5N)_{\gamma}
\nonumber\\&+&
\mathcal{V}_2m_{N}(\rlap/pC)_{\alpha\beta}(\rlap/x\gamma_5N)_{\gamma}+
\mathcal{V}_3m_{N}(\gamma_\mu
C)_{\alpha\beta}(\gamma^\mu\gamma_5N)_{\gamma}+
\mathcal{V}_4m_{N}^2(\rlap/xC)_{\alpha\beta}(\gamma_5N)_{\gamma}\nonumber\\&+&
\mathcal{V}_5m_{N}^2(\gamma_\mu
C)_{\alpha\beta}(i\sigma^{\mu\nu}x_\nu\gamma_5N)_{\gamma} +
\mathcal{V}_6m_{N}^3(\rlap/xC)_{\alpha\beta}(\rlap/x\gamma_5N)_{\gamma}
+(\mathcal{A}_1\nonumber\\
&+&\frac{x^2m_{N}^2}{4}\mathcal{A}_1^M)(\rlap/p\gamma_5
C)_{\alpha\beta}N_{\gamma}+
\mathcal{A}_2m_{N}(\rlap/p\gamma_5C)_{\alpha\beta}(\rlap/xN)_{\gamma}+
\mathcal{A}_3m_{N}(\gamma_\mu\gamma_5 C)_{\alpha\beta}(\gamma^\mu
N)_{\gamma}\nonumber\\&+&
\mathcal{A}_4m_{N}^2(\rlap/x\gamma_5C)_{\alpha\beta}N_{\gamma}+
\mathcal{A}_5m_{N}^2(\gamma_\mu\gamma_5
C)_{\alpha\beta}(i\sigma^{\mu\nu}x_\nu N)_{\gamma}+
\mathcal{A}_6m_{N}^3(\rlap/x\gamma_5C)_{\alpha\beta}(\rlap/x
N)_{\gamma}\nonumber\\&+&(\mathcal{T}_1+\frac{x^2m_{N}^2}{4}\mathcal{T}_1^M)(p^\nu
i\sigma_{\mu\nu}C)_{\alpha\beta}(\gamma^\mu\gamma_5
N)_{\gamma}+\mathcal{T}_2m_{N}(x^\mu p^\nu
i\sigma_{\mu\nu}C)_{\alpha\beta}(\gamma_5 N)_{\gamma}\nonumber\\&+&
\mathcal{T}_3m_{N}(\sigma_{\mu\nu}C)_{\alpha\beta}(\sigma^{\mu\nu}\gamma_5
N)_{\gamma}+
\mathcal{T}_4m_{N}(p^\nu\sigma_{\mu\nu}C)_{\alpha\beta}(\sigma^{\mu\rho}x_\rho\gamma_5
N)_{\gamma}\nonumber\\&+& \mathcal{T}_5m_{N}^2(x^\nu
i\sigma_{\mu\nu}C)_{\alpha\beta}(\gamma^\mu\gamma_5 N)_{\gamma}+
\mathcal{T}_6m_{N}^2(x^\mu p^\nu
i\sigma_{\mu\nu}C)_{\alpha\beta}(\rlap/x\gamma_5
N)_{\gamma}\nonumber\\&+&
\mathcal{T}_7m_{N}^2(\sigma_{\mu\nu}C)_{\alpha\beta}(\sigma^{\mu\nu}\rlap/x\gamma_5
N)_{\gamma}+
\mathcal{T}_8m_{N}^3(x^\nu\sigma_{\mu\nu}C)_{\alpha\beta}(\sigma^{\mu\rho}x_\rho\gamma_5
N)_{\gamma} \, \, ,
\end{eqnarray}
where, the calligraphic objects which  have not definite twists are
functions of the scalar product $px$ and the parameters $a_i$,
$i=1,2,3$ and they are presented in terms of the nucleon
distribution amplitudes (DA's) with definite and  increasing twists.
The scalar, pseudo-scalar, vector, axial vector and tensor DA's are
explicitly shown  in Tables \ref{tab:1}, \ref{tab:2}, \ref{tab:3},
\ref{tab:4} and \ref{tab:5}, respectively.
\begin{table}[h]
\centering
\begin{tabular}{|c|} \hline
$\mathcal{S}_1 = S_1$\\\cline{1-1}\hline
 $2px\mathcal{S}_2=S_1-S_2$ \\\cline{1-1}
   \end{tabular}
\vspace{0.3cm} \caption{Relations between the calligraphic functions
and nucleon scalar DA's.}\label{tab:1}
\end{table}
\begin{table}[h]
\centering
\begin{tabular}{|c|} \hline
  $\mathcal{P}_1=P_1$\\\cline{1-1}
  $2px\mathcal{P}_2=P_1-P_2$ \\\cline{1-1}
   \end{tabular}
\vspace{0.3cm} \caption{Relations between the calligraphic functions
and nucleon pseudo-scalar DA's.}\label{tab:2}
\end{table}
\begin{table}[h]
\centering
\begin{tabular}{|c|} \hline
  $\mathcal{V}_1=V_1$ \\\cline{1-1}
  $2px\mathcal{V}_2=V_1-V_2-V_3$ \\\cline{1-1}
  $2\mathcal{V}_3=V_3$ \\\cline{1-1}
  $4px\mathcal{V}_4=-2V_1+V_3+V_4+2V_5$ \\\cline{1-1}
  $4px\mathcal{V}_5=V_4-V_3$ \\\cline{1-1}
  $4(px)^2\mathcal{V}_6=-V_1+V_2+V_3+V_4
 + V_5-V_6$ \\\cline{1-1}
 \end{tabular}
\vspace{0.3cm} \caption{Relations between the calligraphic functions
and nucleon vector DA's.}\label{tab:3}
\end{table}
\begin{table}[h]
\centering
\begin{tabular}{|c|} \hline
  $\mathcal{A}_1=A_1$ \\\cline{1-1}
  $2px\mathcal{A}_2=-A_1+A_2-A_3$ \\\cline{1-1}
   $2\mathcal{A}_3=A_3$ \\\cline{1-1}
  $4px\mathcal{A}_4=-2A_1-A_3-A_4+2A_5$ \\\cline{1-1}
  $4px\mathcal{A}_5=A_3-A_4$ \\\cline{1-1}
  $4(px)^2\mathcal{A}_6=A_1-A_2+A_3+A_4-A_5+A_6$ \\\cline{1-1}
 \end{tabular}
\vspace{0.3cm} \caption{Relations between the calligraphic functions
and nucleon axial vector DA's.}\label{tab:4}
\end{table}
\begin{table}[h]
\centering
\begin{tabular}{|c|} \hline
  $\mathcal{T}_1=T_1$ \\\cline{1-1}
  $2px\mathcal{T}_2=T_1+T_2-2T_3$ \\\cline{1-1}
   $2\mathcal{T}_3=T_7$ \\\cline{1-1}
  $2px\mathcal{T}_4=T_1-T_2-2T_7$ \\\cline{1-1}
  $2px\mathcal{T}_5=-T_1+T_5+2T_8$ \\\cline{1-1}
  $4(px)^2\mathcal{T}_6=2T_2-2T_3-2T_4+2T_5+2T_7+2T_8$ \\\cline{1-1}
  $4px \mathcal{T}_7=T_7-T_8$\\\cline{1-1}
  $4(px)^2\mathcal{T}_8=-T_1+T_2 +T_5-T_6+2T_7+2T_8$\\\cline{1-1}
 \end{tabular}
\vspace{0.3cm} \caption{Relations between the calligraphic functions
and nucleon tensor DA's.}\label{tab:5}
\end{table}

  The distribution amplitudes $F(a_ipx)$=  $S_i$,
$P_i$, $V_i$, $A_i$, $T_i$ can be written as:
\begin{equation}\label{dependent1}
F(a_ipx)=\int dx_1dx_2dx_3\delta(x_1+x_2+x_3-1) e^{-ip
x\Sigma_ix_ia_i}F(x_i)\; .
\end{equation}
where, $x_{i}$ with $i=1,~2,~3$ corresponds to the longitudinal
momentum fractions carried by the quarks.

In order to obtain the QCD or theoretical representation of the
correlation function, the heavy quark propagator and nucleon
distribution amplitudes are used in Eq. (\ref{mut.m}). Performing
integral over  $x$, equating the corresponding structures from both
representations of the correlation function through the dispersion
relations and applying Borel transformation with respect to
$(p+q)^2$ to suppress the contribution of the higher states and
continuum, one can obtain sum rules for the  form factors $f_{1}$,
$f_{2}$, $f_{3}$, $g_{1}$, $g_{2}$ and $g_{3}$.

By means of the heavy quark effective theory (HQET), the number of
independent form factors  is reduced to two,  $F_1$ and $F_2$.
Hence, the transition matrix element can be parameterized in terms
of these two form factors as \cite{Mannel,alievozpineci}:
\begin{eqnarray}\label{matrixel1111}
\langle N(p)\mid \bar u\Gamma b\mid \Lambda_Q(p+q)\rangle&=&\bar
N(p)[F_1(Q^2)+\not\!vF_2(Q^2)]\Gamma u_{\Lambda_Q}(p+q),\nonumber\\
\end{eqnarray}
where, $\Gamma$ is any Dirac matrices and
$\not\!v=\frac{\not\!p+\not\!q}{m_{\Lambda_{Q}}}$. One can
immediately obtain the  following relations among the form
factors in HQET limit comparing the   Eq. (\ref{matrixel1111}) with
the general definition of the form factors in Eq. (\ref{matrixel1})
(see also  \cite{Chen,ozpineci})
\begin{eqnarray}\label{matrixel22222}
g_1 = f_1=F_{1} + \frac{m_N}{m_{\Sigma_b}}F_{2}\nonumber\\
g_2 = f_2 = g_3 = f_3=\frac{F_2}{m_{\Sigma_b}}
\end{eqnarray}
Considering the above relations, instead of giving the explicit
expressions of the sum rules for the all form factors which are very
lengthy, we will present only the expressions for $f_1$ and $f_2$ in
the Appendix--A. However, we will give the extrapolation of all form
factors in finite mass in terms of $q^2$ in the numerical analysis
section.
\begin{table}[h]
\renewcommand{\arraystretch}{1.5}
\addtolength{\arraycolsep}{3pt}

$$
\begin{array}{|c|r@{}l|r@{}l|r@{}l|r@{ }l|}                                  \hline \hline
  & \multicolumn{4}{c|}{\mbox{HQET }}                    \\
  & \multicolumn{2}{c} {\mbox{QCD sum rulesinput parameters }}
 & \multicolumn{2}{c|}{\mbox{Lattice QCD input parameters}}                            \\ \hline
  \frac{f_{1}}{g_{1}}      &&0 && 0   \\
    \frac{f_{2}}{g_{2}} && $20~\%$ && 0  \\
   \frac{f_{3}}{g_{3}}&& $20~\%$&& 0 \\
   \frac{f_{2}}{g_{3}} &&  $20~\%$&& 0\\
   \frac{f_{3}}{g_{2}}&& $20~\%$ && 0\\
   \frac{f_{2}}{f_{3}} &&0 &&0 \\
   \frac{g_{2}}{g_{3}} && 0 && 0 \\   \hline \hline
\end{array}
$$

\caption{Deviation of the ratio of the form factors from unity (violation of HQET
symmetry relations)  for $\Lambda_{b}\rightarrow p \ell\nu$.}
\label{tab:11}
\renewcommand{\arraystretch}{1}
\addtolength{\arraycolsep}{-1.0pt}
\end{table}

\begin{table}[h]
\renewcommand{\arraystretch}{1.5}
\addtolength{\arraycolsep}{3pt}

$$
\begin{array}{|c|r@{}l|r@{}l|r@{}l|r@{ }l|}                                  \hline \hline
  & \multicolumn{4}{c|}{\mbox{HQET }}                    \\
 & \multicolumn{2}{c} {\mbox{QCD sum rules input parameters}}
 & \multicolumn{2}{c|}{\mbox{Lattice QCD input parameters}}                            \\ \hline
  \frac{f_{1}}{g_{1}}      &&0 && 0   \\
    \frac{f_{2}}{g_{2}}&& $45~\%$ &&$40~\%$  \\
   \frac{f_{3}}{g_{3}}&& $45~\%$&& $40~\%$ \\
   \frac{f_{2}}{g_{3}} &&  $45~\%$&& $35~\%$\\
   \frac{f_{3}}{g_{2}} && $40~\%$ && $35~\%$\\
   \frac{f_{2}}{f_{3}} &&0 &&0 \\
   \frac{g_{2}}{g_{3}} && 0 && 0 \\   \hline \hline
\end{array}
$$
\caption{Deviation of the ratio of the form factors from unity (violation of HQET
symmetry relations) for $\Lambda_{c}\rightarrow n \ell\nu$.}
\label{tab:12}
\renewcommand{\arraystretch}{1}
\addtolength{\arraycolsep}{-1.0pt}
\end{table}

In the following, some remarks about how the HQET limit of the form factors satisfy the above relations are in order. In HQET, all the ratios, $\frac{f_{1}}{g_{1}}$,    $\frac{f_{2}}{g_{2}}$, $ \frac{f_{3}}{g_{3}}$, $\frac{f_{2}}{g_{3}}$, $\frac{f_{3}}{g_{2}}$, $  \frac{f_{2}}{f_{3}}$ and $ \frac{g_{2}}{g_{3}}$ should be equal to one. The deviation of those ratios from unity are presented in  Tables \ref{tab:11} and   \ref{tab:12} for $\Lambda_{b}\rightarrow p\ell\nu$ and $\Lambda_{c}\rightarrow
n\ell\nu$, respectively.  The bottom case and  Lattice QCD input
parameters satisfies the HQET relations exactly, while the maximum violation of this symmetry is related to the charm case and QCD input parameters.  When we consider all relations, we see that the violations for charm case is larger than that of the bottom one.

The explicit expressions of the sum rules for  form factors reveals that to get the numerical values of the form factors, the
expression for residue $\lambda_{\Lambda_{Q}}$ is needed.  This residue
has been calculated in  \cite{Ozpineci1} using two-point QCD sum rules
method:

\begin{eqnarray}\label{residu2}
-\lambda_{\Lambda_{Q}}^{2}e^{-m_{\Lambda_{Q}}^{2}/M_B^{2}}&=&\int_{m_{Q}^{2}}^{s_{0}}e^{\frac{-s}{M_B^{2}}}\rho(s)ds+e^{\frac{-m_Q^2}{M_B^{2}}}\Gamma,
\end{eqnarray}
with
\begin{eqnarray}\label{residurho1}
\rho(s)&=&(<\overline{d}d>+<\overline{u}u>)\frac{(\beta-1)}{192
\pi^{2}} \Bigg\{\frac{m_{0}^{2}}{4
m_{Q}}[6(1+\beta)\psi_{00}-(7+11\beta)\psi_{02}\nonumber\\&-&6(1+\beta)
\psi_{11}]+(1+5\beta)m_{Q}(2\psi_{10}-\psi_{11}-\psi_{12}+2\psi_{21})\Bigg\}\nonumber\\&+&\frac{
m_{Q}^{4}}{2048
\pi^{4}} [5+\beta(2+5\beta)][12\psi_{10}-6\psi_{20}+2\psi_{30}-4\psi_{41}+\psi_{42}-12 ln(\frac{s}{m_{Q}^{2}})],\nonumber\\
\end{eqnarray}

\begin{eqnarray}\label{lamgamma1}
\Gamma&=&\frac{
(\beta-1)}{72}<\overline{d}d><\overline{u}u>\left[\vphantom{\int_0^{x_2}}\right.\frac{m_{Q}^{2}m_{0}^{2}}{2
M_B^{4}}(13+11\beta)+\frac{m_{0}^{2}}{4
M_B^{2}}(25+23\beta)-(13+11\beta)\Bigg].
\end{eqnarray}
where, $s_0$ is continuum threshold, $M_B^2$ is the Borel mass
parameter and $\psi_{nm}=\frac{(s-m_Q^2)^n}{s^m(m_Q^2)^{n-m}}$ are
some dimensionless functions.
\begin{table}[h]
\centering
\begin{tabular}{|c||c|c|} \hline
& QCD sum rules \cite{Lenz} & Lattice QCD
\cite{Gockeler1,Gockeler2,QCDSF}
\\\cline{1-3} \hline\hline
$f_{N}$ & $(5.0\pm0.5)\times10^{-3}~GeV^{2}$ &
$(3.234\pm0.063\pm0.086)\times10^{-3}~GeV^{2}$
\\\cline{1-3} $\lambda_{1}$ &$-(2.7\pm0.9)\times10^{-2}~GeV^{2}$ & $(-3.557\pm0.065\pm0.136)\times10^{-2}~GeV^{2}$ \\\cline{1-3}
 $\lambda_{2}$
&$(5.4\pm1.9)\times10^{-2}~GeV^{2}$&
$(7.002\pm0.128\pm0.268)\times10^{-2}~GeV^{2}$\\\cline{1-3}
$V_{1}^{d}$ &$0.23\pm0.03$& $0.3015\pm0.0032\pm0.0106$
\\\cline{1-3}
$A_{1}^{u}$ &$0.38\pm0.15$& $0.1013\pm0.0081\pm0.0298$\\\cline{1-3}
$f_{1}^{d}$ &$0.40\pm0.05$& $-$\\\cline{1-3} $f_{1}^{u}$
&$0.07\pm0.05$& $-$\\\cline{1-3} $f_{2}^{d}$ &$0.22\pm0.05$&
$-$\\\cline{1-3}
\end{tabular}
\vspace{0.8cm} \caption{The values of independent parameters
entering to the nucleon DA's. The first errors in lattice values are
statistical and the second errors represent the uncertainty due to
the Chiral extrapolation and renormalization.} \label{kazem}
\end{table}
\section{Numerical results}
The numerical analysis of the form factors and  total decay rate for
$\Lambda_{b(c)}\longrightarrow p(n) \ell\nu$ transition are
presented in this section. Some input parameters used in the
analysis of the sum rules for the form factors are $\uu(1~GeV) =
\dd(1~GeV)= -(0.243)^3~GeV^3$, $m_n = 0.939~GeV$, $m_p = 0.938~GeV$,
$m_b = 4.7~GeV$, $m_c = 1.23~GeV$, $m_{\Lambda_{b}} = 5.620~GeV$,
$m_{\Lambda_{c}} = 2.286~GeV$ and $m_0^2(1~GeV) = (0.8\pm0.2)~GeV^2$
\cite{Belyaev}. The main inputs which are the nucleon DA's
can be found in \cite{Lenz}. These DA's contain eight independent
parameters $f_{N},~\lambda_{1},
~\lambda_{2},~V_{1}^{d},~A_{1}^{u},~f_{1}^{d},~f_{1}^{u}$ and
$f_{2}^{d}$. These parameters have been calculated in the light cone
QCD sum rules \cite{Lenz} and also most of these parameters have
been computed in the framework of the lattice QCD
\cite{Gockeler1,Gockeler2,QCDSF}.  For those parameters which have not calculated in lattice, the data from QCD input
parameters will be used. These parameters are given in Table
\ref{kazem}.

Three auxiliary parameters are encountered to the expression ofthe sum rules for form factors,
continuum threshold $s_0$, Borel mass parameter $M_B^2$ and general
parameter $\beta$ entering to the most general form of the interpolating current for  $\Lambda_Q$
baryon. A working region should be
determined for these auxiliary and mathematical parameters such that the form factors as physical quantities
should be independent of them. The continuum threshold, $s_0$ is not
completely arbitrary and it is related to the energy of the exited
states. From our results, we observed that the form factors are
weakly dependent on $s_0$ in the interval,
$(m_{\Lambda_Q}+0.5)^2\leq s_0\leq (m_{\Lambda_Q}+0.7)^2$.  To
determine the working region for  $\beta$, we look at the variation of  the  form
factors with respect to  $ cos\theta$ in the interval $-1\leq
cos\theta\leq1$ which is corresponds to $-\infty\leq
\beta\leq\infty$, where $\beta=tan\theta$. As a result, we
attain a region at which the dependency is weak. The working region
for $\beta$ is obtained to be $-0.75\leq cos\theta\leq0.25$ for
$\Lambda_{b}$ and $-0.25\leq cos\theta\leq0.25$ for $\Lambda_{c}$.
The Ioffe current which corresponds to $cos\theta=-0.71$ is inside the working region for  $\Lambda_{b}$ but out of the region for $\Lambda_{c}$.

For further analysis, the  upper and lower limits of $M_B^{2}$ should be
determined. To do that, we apply two conditions: The first one, which
gives the upper limit, is that the series of the light cone
expansion with increasing twist should be convergent, and the second
one, which determine the lower limit, is that the contribution of higher
states and continuum to the correlation function should be enough
small i.e., the contribution of the highest term with power $1/M_B^2$
is less than, say, 20--25\% of the highest power of $M_B^{2}$. In the present
work, both conditions are satisfied in the region $15 ~GeV^{2}\leq
M_{B}^{2}\leq 30~ GeV^{2}$ for $\Lambda_{b}$ and  $4 ~GeV^{2}\leq
M_{B}^{2}\leq 12~ GeV^{2}$ for $\Lambda_{c}$, which we will use in
numerical analysis. Taking into account the above requirements, we
obtained that the form factors obey  the following extrapolations in
terms of $q^2$:

\begin{equation}\label{17au}
 f_{i}(q^2)[g_{i}(q^2)]=\frac{a}{(1-\frac{q^2}{m^2_{fit}})}+\frac{b}{(1-\frac{q^2}{m^2_{fit}})^2},
\end{equation}
 The values of the parameters
$a,~b$ and $m_{fit}$ for form factors and their HQET limit are given in  Tables \ref{tab:13},
\ref{tab:14}, \ref{tab:31} and \ref{tab:41} related to the  QCD sum
rules and lattice QCD input parameters. Because of the
working near the light cone, $x^2\simeq0$ and concerning the considered correlation function, the results are  not reliable
 at low $q^2$, hence to make the extension of our predictions  to
full physical region, we need to the above parameterization. From those Tables, we see that the pole of the form factors exist outside the physical region and the form factors are analytic in the whole physical interval. The values of form factors at $q^2=0$ obtained from fit functions are  shown in
Tables \ref{tab:15} and \ref{tab:16} for  $\Lambda_{b}\rightarrow p
\ell\nu$ and  $\Lambda_{c}\rightarrow n \ell\nu$, respectively. A comparison of the existing predictions from other approaches is also presented for bottom case. The Table \ref{tab:15} depicts a good consistency on our result for $f_{1}(0)$ HQET limit obtained
from lattice QCD input parameters  with the prediction of
\cite{yeni5}, however the $f_{1}(0)$ HQET limit obtained from QCD sum rules
parameters is almost four times larger than that of
\cite{yeni5} prediction. On the other hand, the similar comparison of our result on form
factor $f_{2}(0)$ at HQET and prediction of  \cite{yeni5} shows that the value presented in
\cite{yeni5} is almost two times greater than our result obtained from Lattice
QCD input parameters and $1.5$ times smaller than our result
obtained from QCD input parameters.
\begin{table}[h]
\renewcommand{\arraystretch}{1.5}
\addtolength{\arraycolsep}{3pt}

$$
\begin{array}{|c|c|c|c|c|c|c|}
\hline \hline
     & \multicolumn{3}{c|}{\mbox{QCD sum rules}} & \multicolumn{3}{c|}{\mbox{Lattice QCD }} \\
\hline
     & \mbox{a} & \mbox{b} & \mbox{$m_{fit}$}    & \mbox{a} & \mbox{b} & \mbox{$m_{fit}$}   \\
\hline
 f_1 &  0.025   &  0.052   & 4.91                &  0.048   &  0.016   & 4.89               \\
 f_2 &  0.007   & -0.050   & 4.92                & -0.003   & -0.006   & 4.92               \\
 f_3 &  0.052   & -0.13    & 4.99                &  0.028   & -0.063   & 4.96               \\
 g_1 & -0.059   &  0.13    & 5.29                & -0.17    &  0.32    & 5.32               \\
 g_2 &  0.011   & -0.050   & 5.20                &  0.019   & -0.040   & 5.40               \\
 g_3 & -0.009   & -0.017   & 4.90                & -0.015   &  0.012   & 4.98               \\
\hline
\hline
\end{array}
$$
\caption{Parameters appearing in  fit function of the original form factors for
$\Lambda_{b}\rightarrow p \ell\nu$.} \label{tab:13}
\renewcommand{\arraystretch}{1}
\addtolength{\arraycolsep}{-1.0pt}
\end{table}

\begin{table}[h]
\renewcommand{\arraystretch}{1.5}
\addtolength{\arraycolsep}{3pt}

$$
\begin{array}{|c|c|c|c|c|c|c|}                                  \hline \hline
 & \multicolumn{3}{c|}{\mbox{QCD sum rules}} & \multicolumn{3}{c|}{\mbox{Lattice QCD }}                             \\
 \hline
         & \mbox{a} & \mbox{b} & \mbox{$m_{fit}$}    & \mbox{a} & \mbox{b} & \mbox{$m_{fit}$}                      \\
\hline
 f_{1}   & -0.034   &  0.20    & 1.59                & -0.14    &  0.64    & 1.55                                  \\
 f_{2}   & -0.015   & -0.77    & 1.57                &  0.018   & -0.32    & 1.60                                  \\
 f_{3}   & -0.062   & -1.23    & 1.48                &  0.12    & -1.09    & 1.56                                  \\
 g_{1}   & -0.015   &  0.54    & 1.53                & -0.20    &  0.71    & 1.59                                  \\
 g_{2}   & -0.11    & -0.20    & 1.52                & -0.034   & -0.14    & 1.65                                  \\
 g_{3}   & -0.088   &  0.085   & 1.48                &  0.009   & -0.41    & 1.50                                  \\
\hline
\hline
\end{array}
$$
\caption{Parameters appearing in   fit function of the  original form factors for
$\Lambda_{c}\rightarrow n \ell\nu$.} \label{tab:14}
\renewcommand{\arraystretch}{1}
\addtolength{\arraycolsep}{-1.0pt}
\end{table}

\begin{table}[h]
\renewcommand{\arraystretch}{1.5}
\addtolength{\arraycolsep}{3pt}

$$
\begin{array}{|c|c|c|c|c|c|c|}
\hline \hline
     & \multicolumn{3}{c|}{\mbox{QCD sum rules}} & \multicolumn{3}{c|}{\mbox{Lattice QCD }} \\
\hline
     & \mbox{a} & \mbox{b} & \mbox{$m_{fit}$}    & \mbox{a} & \mbox{b} & \mbox{$m_{fit}$}   \\
\hline
  f_1 &  0.041   &  0.040   & 4.82                &  0.0042  &  0.016   & 4.92               \\
 f_2 &  0.033   & -0.097   & 4.83                &  0.013   & -0.030   & 5.92               \\
 f_3 &  0.060   & -0.14    & 4.90                &  0.016   & -0.040   & 4.94               \\
 g_1 & -0.0012  &  0.096   & 5.10                & -0.0022  &  0.029   & 5.30               \\
 g_2 & -0.0094  & -0.018   & 5.36                &  0.0017  & -0.0043  & 5.36               \\
 g_3 & -0.040   &  0.025   & 4.95                & -0.018   &  0.015   & 4.98               \\
\hline \hline
\end{array}
$$
\caption{Parameters appearing in the fit function of the form factors at HQET limit  for
$\Lambda_{b}\rightarrow p \ell\nu$.} \label{tab:31}
\renewcommand{\arraystretch}{1}
\addtolength{\arraycolsep}{-1.0pt}
\end{table}
\begin{table}[h]
\renewcommand{\arraystretch}{1.5}
\addtolength{\arraycolsep}{3pt}

$$
\begin{array}{|c|c|c|c|c|c|c|}                                  \hline \hline
 & \multicolumn{3}{c|}{\mbox{QCD sum rules}} & \multicolumn{3}{c|}{\mbox{Lattice QCD }}                             \\
 \hline
         & \mbox{a} & \mbox{b} & \mbox{$m_{fit}$}    & \mbox{a} & \mbox{b} & \mbox{$m_{fit}$}                      \\
\hline
 f_{1}   & -0.066   &  1.14    & 1.51                & -0.039   &  0.37    & 1.55                                  \\
 f_{2}   &  0.046   & -1.14    & 1.53                &  0.047   & -0.63    & 1.48                                  \\
 f_{3}   &  0.071   & -1.33    & 1.50                &  0.039   & -0.52    & 1.53                                  \\
 g_{1}   & -0.10    &  1.21    & 1.57                & -0.039   &  0.39    & 1.55                                  \\
 g_{2}   & -0.070   & -0.11    & 1.56                & -0.027   & -0.046   & 1.60                                  \\
 g_{3}   & -0.076   & -0.91    & 1.54                & -0.034   & -0.032   & 1.54                                  \\
\hline \hline
\end{array}
$$
\caption{Parameters appearing in the fit function of the form factors at HQET limit for
$\Lambda_{c}\rightarrow n \ell\nu$.} \label{tab:41}
\renewcommand{\arraystretch}{1}
\addtolength{\arraycolsep}{-1.0pt}
\end{table}

\begin{table}[h]
\renewcommand{\arraystretch}{1.5}
\addtolength{\arraycolsep}{3pt}

$$
\begin{array}{|c|c|c|c|c|c|}                                  \hline \hline
 & \multicolumn{2}{c|}{\mbox{Original}} & \multicolumn{3}{c|}{\mbox{HQET }}                    \\
 \hline
 & \multicolumn{1}{c|}{\mbox{QCD sum rules}} & \multicolumn{1}{c|}{\mbox{Lattice QCD}}& \multicolumn{1}{c|} {\mbox{QCD sum rules}}
 & \multicolumn{1}{c|} {\mbox{Lattice QCD}}
 & \multicolumn{1}{c|}{\mbox{\cite{yeni5} }}                                       \\ \hline
 f_{1}(0)     & 0.077  &  0.064  &  0.081 &   0.021   &  0.023^{+0.006}_{-0.005}   \\
 f_{2}(0)     &-0.044  & -0.013  & -0.064 &  -0.018   & -0.039^{+0.006}_{-0.009}   \\
 f_{3}(0)     &-0.079  & -0.036  &  -     &   -       &                            \\
 g_{1}(0)     & 0.073  & 0.15    &  -     &   -       &                            \\
 g_{2}(0)     &-0.039  & -0.021  &  -     &   -       &                            \\
 g_{3}(0)     &-0.026  & -0.0035 &  -     &   -       &                            \\
 \hline \hline
\end{array}
$$

\caption{The values of the form factors at $q^2=0$  for $\Lambda_{b}\rightarrow
p \ell\nu$.} \label{tab:15}
\renewcommand{\arraystretch}{1}
\addtolength{\arraycolsep}{-1.0pt}
\end{table}

\begin{table}[h]
\renewcommand{\arraystretch}{1.5}
\addtolength{\arraycolsep}{3pt}

$$
\begin{array}{|c|c|c|c|c|}                                  \hline \hline
 & \multicolumn{2}{c|}{\mbox{Original}} & \multicolumn{2}{c|}{\mbox{HQET }}                    \\
 & \multicolumn{1}{c}{\mbox{QCD sum rules}} & \multicolumn{1}{c|}{\mbox{Lattice QCD}}
 & \multicolumn{1}{c} {\mbox{QCD sum rules}}
 & \multicolumn{1}{c|}{\mbox{Lattice QCD }}                            \\ \hline
 f_{1}(0)     &  0.17   & 0.50   &  1.078  &  0.33   \\
 f_{2}(0)     & -0.78   & -0.31  & -1.09   & -0.58  \\
 f_{3}(0)     &  1.29   & 0.98   &   -     & -      \\
 g_{1}(0)     &  0.52   & 0.51   &   -     & -      \\
 g_{2}(0)     & -0.31   & -0.18  &   -     & -      \\
 g_{3}(0)     & -0.0032 & -0.31  &   -     & -      \\
 \hline \hline
\end{array}
$$

\caption{The values of the form factors  at $q^2=0$ for $\Lambda_{c}\rightarrow
n \ell\nu$.} \label{tab:16}
\renewcommand{\arraystretch}{1}
\addtolength{\arraycolsep}{-1.0pt}
\end{table}
\begin{table}[h] \centering
\tiny{\begin{tabular}{|c||c|c|c|c|c|c|c|} \hline &
$\Lambda_{b}\longrightarrow p \mu\nu_{\mu}$
&$\Lambda_{b}\longrightarrow p e\nu_{e}$
&$\Lambda_{b}\longrightarrow p \tau\nu_{\tau}$&
$\Lambda_{c}\longrightarrow n \mu\nu_{\mu}$ &
$\Lambda_{c}\longrightarrow n e\nu_{e}$ &$\Lambda_{c}\longrightarrow
n \tau\nu_{\tau}$
\\\cline{1-6}\hline\hline
For QCD sum rules inputs& $(3.07\pm1.05) \times 10^{-15}$
&$(3.065\pm1.05)\times 10^{-15}$& $(3.82\pm1.35) \times 10^{-15}$&
$(2.89\pm0.95)\times 10^{-13}$& $(2.86\pm0.95) \times 10^{-13}$& -\\
\cline{1-7} For lattice QCD inputs& $(2.87\pm0.95) \times 10^{-15}$ &
$(2.87\pm0.95) \times 10^{-15}$& $(2.55\pm0.85) \times 10^{-15}$&
$(1.35\pm0.45) \times 10^{-13}$& $(1.33\pm0.43) \times 10^{-13}$&
-\\\cline{1-7}HQET limit for QCD sum rules inputs& $(5.84\pm1.81) \times
10^{-15}$ &$(5.83\pm1.81)\times 10^{-15}$& $(7.90\pm2.45) \times
10^{-15}$&
$(5.08\pm1.65)\times 10^{-13}$& $(5.01\pm1.60) \times 10^{-13}$& -\\
\cline{1-7} HQET limit for lattice QCD inputs& $(4.70\pm1.60) \times
10^{-17}$ & $(4.60\pm1.55) \times 10^{-17}$& $(2.36\pm0.85) \times
10^{-16}$& $(8.75\pm2.85) \times 10^{-14}$& $(8.74\pm2.83) \times
10^{-14}$& -\\\cline{1-6}\hline\hline
\end{tabular}
\vspace{0.8cm} \caption{Values of the
$\Gamma(\Lambda_{Q}\longrightarrow N \ell\nu$) in  GeV  for
different leptons and two sets of input parameters obtained
 from QCD sum rules and lattice QCD and also their HQET limit.} \label{tab:27}}
\end{table}
\begin{table}[h]
\renewcommand{\arraystretch}{1.5}
\addtolength{\arraycolsep}{3pt}

$$
\tiny{\begin{array}{|c|c|c|c|c|c|c|}
\hline \hline
 & \multicolumn{1}{c}{\mbox{$\Lambda_{b}\longrightarrow p \mu\nu_{\mu}$}}
 & \multicolumn{1}{c}{\mbox{$\Lambda_{b}\longrightarrow p e\nu_{e}$ }}
 & \multicolumn{1}{c|}{\mbox{$\Lambda_{b}\longrightarrow p \tau\nu_{\tau}$}}
 & \multicolumn{1}{c}{\mbox{$\Lambda_{c}\longrightarrow n \mu\nu_{\mu}$}}
 & \multicolumn{1}{c} {\mbox{$\Lambda_{c}\longrightarrow n e\nu_{e}$}}
 & \multicolumn{1}{c|} {\mbox{$\Lambda_{c}\longrightarrow n \tau\nu_{\tau}$}}\\ \hline
 \mbox {For QCD sum rules}         & (2.5\pm0.85) \times 10^{14}  & (2.5\pm0.85) \times 10^{14}   & (3.12\pm1.05) \times 10^{14}  &
  (8.3\pm2.85) \times 10^{12}      & (8.21\pm2.80) \times 10^{12} &                   \\
\hline
 \mbox{For lattice QCD }           &(2.35\pm0.85) \times10^{14}   &(2.35\pm0.85) \times 10^{14}    & (2.08\pm0.70) \times10^{14}  &
 (3.88\pm1.25) \times 10^{12}      &(3.82\pm1.20) \times10^{12}   &                    \\ \hline
  \mbox {HQET limit for QCD sum rules}         & (4.78\pm1.75) \times 10^{14}  & (4.77\pm1.75) \times 10^{14}   & (6.46\pm2.15) \times 10^{14}  &
  (1.46\pm0.55) \times 10^{13}      & (1.44\pm0.55) \times 10^{13} &                   \\
\hline
 \mbox{HQET limit for lattice QCD }           &(3.84\pm1.25) \times10^{12}   &(3.76\pm1.20) \times 10^{12}    & (1.93\pm0.70) \times10^{12}  &
 (2.51\pm0.85) \times 10^{12}      &(2.51\pm0.85) \times10^{12}   &                    \\ \hline
 \mbox{ \cite{yeni2}}              & \multicolumn{3}{c|}{2.05 \times10^{13}}         & \multicolumn{3}{c|} {}        \\
 \hline
 \mbox{\cite{yeni1}}               &\multicolumn{3}{c|}{2.58\times 10^{13}}          & \multicolumn{3}{c|} {}        \\
 \hline
 \mbox{\cite{Datta}}               &\multicolumn{3}{c|}{6.48 \times10^{12}}          & \multicolumn{3}{c|} {}        \\
 \hline
 \mbox{QCD sum rules\cite{yeni4}}  &\multicolumn{3}{c|}{3.65 \times10^{13}}          & \multicolumn{3}{c|} {}        \\
 \hline
 \mbox{ HQET\cite{yeni4}}          &\multicolumn{3}{c|}{5.62 \times10^{12}}          & \multicolumn{3}{c|} {}        \\
 \hline
 \mbox{\cite{pervin}}              &\multicolumn{2}{c}{4.55 \times 10^{12} \mbox{(HONR)}} &\multicolumn{1}{c|}{4.01 \times 10^{12} \mbox{(HONR)}}
 & \multicolumn{2}{c} {1.02 \times 10^{10}\mbox{(HONR)}} & \multicolumn{1}{c|} {}         \\
 &\multicolumn{2}{c}{7.55 \times 10^{12}\mbox{(HOSR)}}   &\multicolumn{1}{c|}{6.55 \times 10^{12}\mbox{(HOSR)}}
 & \multicolumn{2}{c} {1.35 \times 10^{10}\mbox{(HOSR)}} & \multicolumn{1}{c|} {}          \\
 \hline \hline
\end{array}}
$$

\caption{Values of the total decay rate (in $|V_{qQ}|^{2}~s^{-1}$)
 of the $\Lambda_{Q}\longrightarrow N \ell\nu$ transition for  different leptons and two sets of input parameters obtained
 from QCD sum rules and lattice QCD and also their HQET limit compared to the \cite{Datta,
yeni1,yeni2,yeni4,pervin,yeni5}.} \label{tab:10}
\renewcommand{\arraystretch}{1}
\addtolength{\arraycolsep}{-1.0pt}
\end{table}

 In the next step, we calculate the total decay rate of $\Lambda_{Q}\longrightarrow N \ell\nu$ transition in the whole
physical region, i.e., $ m_{l}^2 \leq q^2 \leq (m_{\Lambda_{Q}} -
m_{N})^2$. The decay width  for such transition is given by the
following expression~\cite{Faessler,Pietschmann:1974ap}
\eq\label{Gamma_BiBf} \Gamma(\Lambda_{Q}\to N l \nu_l) =
\frac{G_F^2}{384 \pi^3 m_{\Lambda_{Q}}^3} \ |V_{\rm qQ}|^2 \,  \,
\int\limits_{m_l^2}^{\Delta^2} dq^2  \ (1 - m_l^2/q^2)^2 \
\sqrt{(\Sigma^2 - q^2) (\Delta^2 - q^2)} \ N(q^2) \en where \eq
N(q^2) &=& F_1^2(q^2) (\Delta^2 (4q^2 - m_l^2) + 2 \Sigma^2 \Delta^2
(1 + 2 m_l^2/q^2) - (\Sigma^2 + 2q^2) (2q^2 + m_l^2) )\nonumber\\[3mm]
&+& F_2^2(q^2) (\Delta^2 - q^2)(2 \Sigma^2 + q^2) (2q^2 +
m_l^2)/m_{\Sigma_b}^2
+ 3 F_3^2(q^2) m_l^2 (\Sigma^2 - q^2) q^2/m_{\Sigma_b}^2  \nonumber\\[3mm]
&+& 6 F_1(q^2) F_2(q^2) (\Delta^2 - q^2) (2 q^2 + m_l^2)
\Sigma/m_{\Sigma_{b}}
- 6 F_1(q^2) F_3(q^2)  m_l^2 (\Sigma^2 - q^2) \Delta/m_{\Sigma_b} \nonumber\\[3mm]
&+& G_1^2(q^2) (\Sigma^2 (4q^2 - m_l^2) + 2 \Sigma^2 \Delta^2
(1 + 2 m_l^2/q^2) - (\Delta^2 + 2q^2) (2q^2 + m_l^2) )\nonumber\\[3mm]
&+& G_2^2(q^2) (\Sigma^2 - q^2)(2 \Delta^2 + q^2) (2q^2 +
m_l^2)/m_{\Sigma_b}^2
+ 3 G_3^2(q^2) m_l^2 (\Delta^2 - q^2) q^2/m_{\Sigma_b}^2  \nonumber\\[3mm]
&-& 6 G_1(q^2) G_2(q^2) (\Sigma^2 - q^2) (2 q^2 + m_l^2)
\Delta/m_{\Sigma_b}
 +  6 G_1(q^2) G_3(q^2)  m_l^2 (\Delta^2 - q^2) \Sigma/m_{\Sigma_b}  \,.
\en
Here, $F_1(q^2)=f_1(q^2)$, $F_2(q^2)=m_{\Lambda_{Q}}f_2(q^2)$, $F_3(q^2)=m_{\Lambda_{Q}}f_3(q^2)$, $G_1(q^2)=g_1(q^2)$, $G_2(q^2)=m_{\Lambda_{Q}}g_2(q^2)$,
 $G_3(q^2)=m_{\Lambda_{Q}}g_3(q^2)$, $\Sigma  =
m_{\Lambda_{Q}} + m_{N}$ and $\Delta  = m_{\Lambda_{Q}} - m_{N}$.
$G_F = 1.17 \times 10^{-5}$ GeV$^{-2}$ is the Fermi coupling
constant, and $m_l$ is the leptonic (electron,  muon or tau) mass.
For the corresponding CKM matrix element
$V_{ub}=(4.31\pm0.30)~10^{-3}$ and $V_{cd}=(0.230\pm0.011)$ are
used~\cite{Yao:2006px}.

 Our final results for total decay rates are
given in Table \ref{tab:27}. As it can be seen from this Table, our
results for $e$ and $\mu$ and $\Lambda_{b}$ cases are consistent for two sets of input parameters when the original form factors are used especially, when we consider the uncertainties.
 However, QCD input parameters result is 1.5 times
greater than that of the lattice input parameter for the decay rates
of $\tau$ and bottom case. If we consider $\Lambda_{c}$, QCD sum rules input
parameters gives the result 2 times greater  than the Lattice
QCD input parameters. On the other hand, when we consider the
uncertainties, results obtained using both sets of input parameters and original form factors
coincide for all leptons. At HQET limit and QCD sum rules input parameters, our predictions for the decay rates are in the same order of magnitude with the original form factors and two sets for all leptons and both charm and bottom cases. In contrast,   the results at HQET limit and lattice parameters are two orders of magnitude less than HQET limit and sum rules inputs as well as original form factors for bottom and $e$ and $\mu$ cases. For $\tau$  and bottom, and $e$ and $\mu$ and charm cases, this difference is approximately one order of magnitude. We also compare our results on decay rates
in units of $|V_{qQ}|^{2}~s^{-1}$ with the predictions
of references \cite{Datta, yeni1,yeni2,yeni4,pervin,yeni5} in Table \ref{tab:10}.  From this Table, it is clear that our results for bottom case, lattice parameters and HQET limit are in the same order of magnitude with the predictions of  \cite{Datta,pervin}
and HQET-\cite{yeni4}. For all other cases the difference between our results with the existing predictions of the other approaches presented in Table \ref{tab:10} is one-two order of magnitudes.  In Table \ref{tab:10},
HOSR refers to harmonic oscillator semi relativistic and HONR stands
for harmonic oscillator non relativistic constituent quark models.

To summarize, using the most general form of the interpolating currents of $\Lambda_{Q}$ and nucleon DA's with two sets of input parameters, namely QCD sum rules and lattice QCD inputs, the transition form factors of the semileptonic
$\Lambda_{Q}\rightarrow Nl\nu$ have been  calculated in the framework of
the light cone QCD sum rules in full theory and HQET. The lattice input parameters satisfy the HQET relations exactly for bottom case, while the maximum violation is for charm case and QCD input parameters. The results of the form factors at HQET and $q^2=0$ have been compared with the existing predictions of the other approaches.  These transition form factors have been used to estimate the corresponding tree level semileptonic decay rates both in full theory and HQET limit. A comparison of the obtained results and the existing predictions of the other approaches which all are at HQET limit, was also presented. The best consistency between our results and those predictions is related to the bottom case and lattice QCD input parameters at HQET. Our results can be checked in experiments hold in future such as LHC. Comparison between the experimental data and our results could give essential information about the nature of the $\Lambda_{Q}$, Nucleon distribution amplitudes as well as determination of the CKM matrix elements, $V_{ub}$ and $V_{cd}$.

 \newpage

\section*{Appendix A}
In this Appendix,  the explicit expressions for the form
factors $f_1$ and $f_2$ are given:
\begin{eqnarray}\label{f_{1}}
&&f_{1}(Q^{2})=\frac{1}{2\sqrt{\lambda_{\Lambda_b}}}
e^{m_{\Lambda_b}^{2}/M_{B}^{2}}\left\{\vphantom{\int_0^{x_2}}
\int_{t_{0}}^{1}dx_{2}\int_{0}^{1-x_{2}}dx_{1}
e^{-s(x_{2},Q^{2})/M_{B}^{2}}\frac{1}{2\sqrt{6} x_{2}}
\left[\vphantom{\int_0^{x_2}}m_b(-1+\beta){\cal
H}_{+11,+17_2,+5}(x_i) \right. \right.\nonumber\\&& \left. \left.-
m_N x_{2} \left(\vphantom{\int_0^{x_2}}{\cal
H}_{-11,-13_3,+17_{10},-19_{18},+3_2,+5_3,-7_5}(x_i) +\beta{\cal
H}_{+11,+13_3,+17_8,-19_{28},+3_4,-5_3,+7_5}(x_i)+2{\cal
H}_{1}(2+\beta) \vphantom{\int_0^{x_2}} \right)
\vphantom{\int_0^{x_2}} \right] \vphantom{\int_0^{x_2}}
\right.\nonumber\\&& \left.
+\int_{t_0}^1dx_2\int_0^{1-x_2}dx_1\int_{t_0}^{x_2}dt_1
e^{-s(t_{1},Q^{2})/M_{B}^{2}} \left[\vphantom{\int_0^{x_2}}
\frac{1}{12 \sqrt{6} M_{B}^{4}
t_{1}^{2}}\left\{\vphantom{\int_0^{x_2}}m_N^{5} x_{2}
\left(\vphantom{\int_0^{x_2}}{\cal
H}_{+10_3,+16,-22_{8},+24_{10}}(x_i)
\right.\right.\right.\right.\nonumber\\&& \left.\left.\left.\left.
-\beta {\cal H}_{+10_3,+16,+22_{4},-24_{8}}(x_i)
\vphantom{\int_0^{x_2}}\right)+ m_N^{4} m_b {\cal
H}_{22}(x_i)(1-\beta)(2+3x_{2})-m_N^{2} m_b x_{2} {\cal
H}_{22}(x_i)(1-\beta)(Q^2+s(t_1, Q^2))
\vphantom{\int_0^{x_2}}\right\}
 \right.\right.\nonumber\\&&
\left.\left.+ \frac{1}{12 \sqrt{6} M_{B}^{4} t_{1}}
\left\{\vphantom{\int_0^{x_2}}m_N^{5}\left(\vphantom{\int_0^{x_2}}
{\cal H}_{+10_3,+16}(x_i)(-1+\beta)(1+x_{2})-2{\cal
H}_{24}(x_i)(5+4\beta)(1+x_{2})+4{\cal
H}_{22}(x_i)(2+\beta)(1+2x_{2}) \vphantom{\int_0^{x_2}}\right)
\right.\right.\right.\nonumber\\&& \left.\left.\left. +m_N^{4}m_b
(-1+\beta)\left(\vphantom{\int_0^{x_2}}x_2{\cal
H}_{-10,+16,-24_2}(x_i)+ {\cal H}_{22}(x_i)(3+x_{2})
\right)-m_N^{3}x_{2}(Q^2+s(t_1,
Q^2))\left(\vphantom{\int_0^{x_2}}{\cal H}_{+22_8,-24_{10}}(x_i)
\right.\right.\right.\right.\nonumber\\&& \left.\left.\left.\left.
+\beta {\cal H}_{+22_4,-24_8}(x_i)+(-1+\beta){\cal
H}_{+10_3,+16}(x_i) \vphantom{\int_0^{x_2}}\right)-m_N^{2}m_b {\cal
H}_{22}(x_i)(-1+\beta)\left(\vphantom{\int_0^{x_2}}Q^2+s(t_1,
Q^2)+3Q^2x_{2}+s(t_1,
Q^2)x_{2}\right)\vphantom{\int_0^{x_2}}\right\}
\right.\right.\nonumber\\&& \left.\left. +\frac{1}{24 \sqrt{6}
M_{B}^2 t_{1}}\left\{\vphantom{\int_0^{x_2}}m_N^{3}
\left(\vphantom{\int_0^{x_2}}{\cal
H}_{-12_{10},+18,-20_{23},-6_6}(x_i)+\beta{\cal
H}_{+12_{10},-18_7,-20_{85},+6_6}(x_i)+x_2{\cal
H}_{+10_{6},+16_2,+24_{20}}(x_i)
\right.\right.\right.\right.\nonumber\\&& \left.\left.\left.\left.
+\beta x_2{\cal H}_{-10_{6},-16_2,+24_{16}}(x_i)
\vphantom{\int_0^{x_2}}\right)-m_N^{2} m_b
(-1+\beta)\left(\vphantom{\int_0^{x_2}}{\cal
H}_{+12_{2},+14,+15,-20_{4},+21_4,-6_2,+8,+9}(x_i)-4x_2{\cal
H}_{22}(x_i)\vphantom{\int_0^{x_2}}\right)
\right.\right.\right.\nonumber\\&& \left.\left.\left. +m_N
(Q^2+s(t_1, Q^2))\left(\vphantom{\int_0^{x_2}}{\cal
H}_{-18_{3},-20_3,-6_{2}}(x_i)+\beta{\cal
H}_{+18,+20_{19},+6_{2}}(x_i)-2 {\cal
H}_{12}(x_i)(-1+\beta)\vphantom{\int_0^{x_2}}\right)
\vphantom{\int_0^{x_2}}\right\} \right.\right.\nonumber\\&&
\left.\left.+\frac{m_N^{4}m_b}{\sqrt{6} M_{B}^4 t_{1}^3}x_2{\cal
H}_{22}(x_i)(-1+\beta) -\frac{m_N^{3}}{2\sqrt{6} M_{B}^2
t_{1}^2}\left(\vphantom{\int_0^{x_2}}{\cal
H}_{+18,+20,+6}(x_i)-\beta{\cal H}_{+18,+20_{11},+6}(x_i)+{\cal
H}_{12}(x_i)(-1+\beta)\vphantom{\int_0^{x_2}}\right)
\right.\right.\nonumber\\&& \left.\left. +\frac{1}{12\sqrt{6}
M_{B}^4}\left\{\vphantom{\int_0^{x_2}}m_N^{5}\left(\vphantom{\int_0^{x_2}}
{\cal H}_{+16,+10_3,+24_{10}}(x_i)+\beta {\cal
H}_{-10_3,-16,+24_{8},}(x_i)+2{\cal
H}_{22}(x_i)(2+\beta)(-4+t_1-x_2)\vphantom{\int_0^{x_2}}\right)
\right.\right.\right.\nonumber\\&& \left.\left.\left.
+m_N^{4}m_b(-1+\beta){\cal
H}_{+10,-16,-22,+24_2}(x_i)+m_N^{3}\left[\vphantom{\int_0^{x_2}}
(-1+\beta)\left(\vphantom{\int_0^{x_2}}Q^2+s(t_1, Q^2)-Q^2 t_1+Q^2
x_2\vphantom{\int_0^{x_2}}\right){\cal H}_{+10_3,+16}(x_i)
\right.\right.\right.\right.\nonumber\\&& \left.\left.\left.\left.
-2{\cal
H}_{22}(x_i)(2+\beta)\left(\vphantom{\int_0^{x_2}}Q^2(-2+3t_1-3
x_2)+s(t_1, Q^2)(-2+t_1-x_2)\vphantom{\int_0^{x_2}}\right)-2{\cal
H}_{24}(x_i)(5+4 \beta)\left(\vphantom{\int_0^{x_2}}Q^2+s(t_1,
Q^2)\right.\right.\right.\right.\right.\nonumber\\&&
\left.\left.\left.\left.\left.-Q^2 t_1+Q^2
x_2\vphantom{\int_0^{x_2}}\right)
\vphantom{\int_0^{x_2}}\right]+m_N^{2}m_b {\cal
H}_{22}(x_i)(-1+\beta) \left(\vphantom{\int_0^{x_2}}3Q^2+s(t_1,
Q^2)\vphantom{\int_0^{x_2}}\right) \vphantom{\int_0^{x_2}}\right\}
+\frac{1}{24\sqrt{6} M_{B}^2}\left\{\vphantom{\int_0^{x_2}}
-4m_N^{2} m_b {\cal H}_{22}(x_i)(-1+\beta)
\right.\right.\right.\nonumber\\&& \left.\left.\left.+ m_N^{3}
\left(\vphantom{\int_0^{x_2}}{\cal
H}_{+12_8,-14,+15_5,-16_2,+18_5,-2_8,+20_{21}+21_{20},+23_{36}
,-24_{20},+4_4,+6_8,+8_3 ,-9_7}(x_i)+8{\cal H}_{22}(x_i)((2+\beta)
t_1+(2+\beta)x_2) \right.\right.\right.\right.\nonumber\\&&
\left.\left.\left.\left. +6{\cal H}_{10}(x_i)(-1+\beta) + \beta
{\cal H}_{-12_8,+14,-15_5,+16_2,+18_7,-2_4,+20_{61}+21_{16},+23_{56}
,-24_{16},+4_8,-6_8, -8_3 ,+9_7}(x_i)
\vphantom{\int_0^{x_2}}\right)\right.\right.\right.\nonumber\\&&
\left.\left.\left. +m_N \left[\vphantom{\int_0^{x_2}} Q^2{\cal
H}_{-12_6,+18,-20_{11},-6_2}(x_i)-s(t_1, Q^2){\cal
H}_{+12_4,+18,+20_{13},+6_4}(x_i)+\beta
\left(\vphantom{\int_0^{x_2}}Q^2 {\cal
H}_{+12_6,-18_5,-20_{55},+6_2}(x_i)
\right.\right.\right.\right.\right.\nonumber\\&&
\left.\left.\left.\left.\left. +s(t_1, Q^2){\cal
H}_{+12_4,-18_3,-20_{33},+6_4}(x_i)\vphantom{\int_0^{x_2}}\right)
\vphantom{\int_0^{x_2}}\right]\vphantom{\int_0^{x_2}}\right\}
-\frac{m_N}{4\sqrt{6}t_1}\left(\vphantom{\int_0^{x_2}}{\cal
H}_{+12_2,+18_3,+20_{15},+6_6}(x_i)+\beta {\cal
H}_{-12_2,+18,+20_{11},-6_6}(x_i)\vphantom{\int_0^{x_2}}\right)
\right.\right.\nonumber\\&& \left.\left.
+\frac{m_N}{4\sqrt{6}}\left(\vphantom{\int_0^{x_2}}{\cal
H}_{+12_4,+18,+20_{13},+6_4}(x_i)+\beta {\cal
H}_{-12_4,+18_3,+20_{33},-6_4}(x_i)\vphantom{\int_0^{x_2}}\right)
\vphantom{\int_0^{x_2}}\right] \right.\nonumber\\&&
\left.+\int_{t_{0}}^{1}dx_{2}\int_{0}^{1-x_{2}}dx_{1}
e^{-s_0/M_{B}^{2}}\left[\vphantom{\int_0^{x_2}}
\frac{1}{(Q^2+m_N^2t_0^2)^3
24\sqrt{6}}\left\{\vphantom{\int_0^{x_2}}-\frac{2 m_N^9
t_0^4}{M_{B}^{2}}(t_0-x_2)\left[\vphantom{\int_0^{x_2}}
(-1+\beta)(-1+t_0){\cal H}_{+10_3,+16}(x_i)
\right.\right.\right.\right.\nonumber\\&&\left.\left. \left.\left.
-2{\cal H}_{24}(x_i)(5+4\beta)(-1+t_0)-2{\cal
H}_{22}(x_i)(2+\beta)\left(\vphantom{\int_0^{x_2}}2-4t_0+t_0^2
\vphantom{\int_0^{x_2}}\right) \vphantom{\int_0^{x_2}}\right]
-\frac{2 m_N^8 m_b
t_0^3}{M_{B}^{2}}(-1+\beta)(t_0-x_2)\right.\right.\right.
\nonumber
\end{eqnarray}
\begin{eqnarray}
&&
 \left.\left.\left.\left(\vphantom{\int_0^{x_2}}{\cal
H}_{22}(x_i)(2-3t_0) +t_0^2{\cal
H}_{-10,+16,+22,-24_2}(x_i)\vphantom{\int_0^{x_2}}\right)+ m_N^7
\left[\vphantom{\int_0^{x_2}}
\frac{2t_0^7}{M_{B}^{2}}\left\{\vphantom{\int_0^{x_2}}Q^2
\left(\vphantom{\int_0^{x_2}} {\cal
H}_{+10_3,+16,+24_{10}}(x_i)-\beta {\cal H}_{+10_3,+16,-24_{8}}(x_i)
\vphantom{\int_0^{x_2}}\right)
\right.\right.\right.\right.\right.\nonumber\\&&
\left.\left.\left.\left.\left. +2{\cal
H}_{22}(x_i)(2+\beta)\left(\vphantom{\int_0^{x_2}}2M_{B}^{2}-3Q^2-s(s_0,
Q^2) \vphantom{\int_0^{x_2}}\right)\vphantom{\int_0^{x_2}}\right\}
 +t_0^6\left\{\vphantom{\int_0^{x_2}}(-1+\beta){\cal H}_{+10_6,-12_8,
 +14,-15_5,+16_2,-6_8,-8_3,+9_7}(x_i)
 \right.\right.\right.\right.\right.\nonumber\\&& \left.\left.\left.
 \left.\left.
 +{\cal H}_{+18_5,-2_8,+20_{21},+21_{20},+22_{24},+23_{36},-24_{20},
 +4_4}(x_i)+
 \beta{\cal H}_{+18_7,-2_4,+20_{61},+21_{16},+22_{12},+23_{56},
 -24_{16},+4_8}(x_i)
\right.\right.\right.\right.\right.\nonumber\\
&&+\left.\left.\left.\left.\left.\frac{1}{M_{B}^{2}}\left[\vphantom{\int_0^{x_2}}
Q^2\left(\vphantom{\int_0^{x_2}}  {\cal
H}_{-10_6,-16_2,+22_{16},-24_{20}}(x_i)+\beta{\cal
H}_{+10_6,+16_2,+22_{8},-24_{16}}(x_i)+x_2{\cal
H}_{-10_6,-16_2,+22_{24},-24_{20}}(x_i)
\right.\right.\right.\right.\right.\right.\right.\nonumber\\
&&+\beta x_2{\cal H}_{+10_6,+16_2,+22_{12},-24_{16}}(x_i)
\left.\vphantom{\int_0^{x_2}}\right)
+s(s_0,Q^2)\left(\vphantom{\int_0^{x_2}} {\cal
H}_{-10_6,-16_2,+22_{16},-24_{20}}(x_i)+\beta {\cal
H}_{+10_6,+16_2,+22_{8},-24_{16}}(x_i) \right.\nonumber\\&& \left.
 +4x_2{\cal
H}_{22}(x_i)(2+\beta)\left.\vphantom{\int_0^{x_2}}\right)
\left.\vphantom{\int_0^{x_2}}\right] \left.
\vphantom{\int_0^{x_2}}\right\}+t_0^5\left\{\vphantom{\int_0^{x_2}}
{\cal
H}_{+10_{18},-12_{10},+16_6,+18,-20_{23},-22_{96},+24_{60},+6_6}(x_i)+x_2
{\cal H}_{+10_6,+16_2,-22_{24},+24_{20}}(x_i)
\right.\right.\nonumber\\&& \left.\left. + \beta{\cal
H}_{-10_{18},+12_{10},-16_6,-18_7,-20_{85},-22_{48},+24_{48},+6_6}(x_i)
+\beta x_2 {\cal
H}_{+10_6,-16_2,-22_{12},+24_{16}}(x_i)+\frac{1}{M_{B}^{2}}
\left[\vphantom{\int_0^{x_2}}8Q^2{\cal H}_{22}(x_i)(2+\beta)
\right.\right.\right.\nonumber\\&& \left.\left.\left.+Q^2
x_2\left(\vphantom{\int_0^{x_2}}{\cal
H}_{+10_6,+16_2,-22_{16},+24_{20}}(x_i)+\beta {\cal
H}_{-10_6,-16_2,-22_{8},+24_{16}}(x_i)
\vphantom{\int_0^{x_2}}\right)+x_2
s(s_0,Q^2)\left(\vphantom{\int_0^{x_2}}{\cal
H}_{+10_6,+16_2,-22_{16},+24_{20}}(x_i)
\right.\right.\right.\right.\nonumber\\&&  \left.\left.\left.
\left.-\beta{\cal
H}_{+10_6,+16_2,+22_{8},+24_{16}}(x_i)\vphantom{\int_0^{x_2}}\right)
\vphantom{\int_0^{x_2}}\right]
\vphantom{\int_0^{x_2}}\right\}+\frac{2
t_0^4}{M_{B}^{2}}\left\{\vphantom{\int_0^{x_2}}M_{B}^{2}{\cal
H}_{-16_3,-18,-20,+22_{24},-24_{30},-6}(x_i)+Q^2 \beta{\cal
H}_{-16_2,-22_{16},+24_{16}}(x_i) \right.\right.\nonumber\\&&
\left.\left. +M_{B}^{2}\beta{\cal
H}_{+16_3,+18,+20_{11},+22_{12},-24_{24},+6}(x_i)+Q^2{\cal
H}_{+16_2,-22_{32},+24_{20}}(x_i)+M_{B}^2{\cal H}_{12}(x_i)(1-\beta)
+ x_2\left(\vphantom{\int_0^{x_2}}-3M_{B}^{2}{\cal H}_{16}(x_i)
\right.\right.\right.\nonumber\\&&
\left.\left.\left.(1-\beta)+4{\cal
H}_{22}(x_i)(2+\beta)(6^{2}-Q^2)-6M_{B}^{2}{\cal
H}_{24}(x_i)(5+4\beta)\vphantom{\int_0^{x_2}}\right)-3{\cal
H}_{10}(x_i)(1-\beta)
(-2Q^2+3M_{B}^{2}(1+x_2))\vphantom{\int_0^{x_2}}\right\}
\right.\nonumber\\&& \left.
 +\frac{2 t_0^3}{M_{B}^{2}}\left\{\vphantom{\int_0^{x_2}}
 \left(\vphantom{\int_0^{x_2}} {\cal
H}_{+16,-10_3}(x_i)(1-\beta)+  2{\cal
H}_{24}(x_i)(5+4\beta)\vphantom{\int_0^{x_2}}\right)(3M_{B}^{2}
x_2-2Q^2(1+x_2))-4{\cal H}_{22}(x_i)(2+\beta)
\right.\right.\nonumber\\&& \left.\left.
  (3M_{B}^{2}
x_2-2Q^2(1+2x_2))\vphantom{\int_0^{x_2}}\right\}-\frac{4 Q^2 t_0^2
x_2}{M_{B}^{2}}\left(\vphantom{\int_0^{x_2}}{\cal
H}_{+22_8,-24_{10}}(x_i)-{\cal
H}_{+10_3,+16}(x_i)(1-\beta)+\beta{\cal
H}_{+22_4,-24_8}(x_i)\vphantom{\int_0^{x_2}}\right) \left.
\vphantom{\int_0^{x_2}}\right] \right.\nonumber\\&& \left.\left.
+\frac{m_N^6 m_b(1-\beta)t_0}{M_{B}^2}\left\{\vphantom{\int_0^{x_2}}
-2{\cal H}_{22}(x_i)(t_0-x_2)\left[\vphantom{\int_0^{x_2}}-M_{B}^2
t_0\left(\vphantom{\int_0^{x_2}}6+t_0(3+t_0)(-3+2t_0)
\vphantom{\int_0^{x_2}}\right) +Q^2\left(\vphantom{\int_0^{x_2}}
-4+t_0(6+t_0(-1+t_0) \right.\right.\right.\right.\right.
\nonumber\\&& \left.\left.\left.\left.\left.
(2+3t_0))\vphantom{\int_0^{x_2}}\right)+(-1+t_0)t_0^3s(s_0, Q^2)
\vphantom{\int_0^{x_2}}\right]-4t_0^3Q^2{\cal
H}_{+10,-16,+24_2}(x_i)+2x_2t_0^2 {\cal
H}_{+10,-16,+24_2}(x_i)(2Q^2+3 M_{B}^2 t_0) \right.\right.\right.
\nonumber\\&& \left.\left. \left.+t_0^4M_{B}^2{\cal
H}_{-10_6,+12_2,+14,+15,+16_6,-20_4,+21_4,-24_{12},-6_2,+8,+9}(x_i)
\vphantom{\int_0^{x_2}}\right\}
+\frac{m_N^5}{M_{B}^2}\left\{\vphantom{\int_0^{x_2}} Q^2
t_0\left(\vphantom{\int_0^{x_2}}{\cal
H}_{-16_2,+22_{16},-24_{20}}(x_i) \right.\right.\right.\right.
\nonumber\\&& \left.\left.\left.\left.+\beta {\cal
H}_{+16_2,+22_{8},-24_{16}}(x_i) \vphantom{\int_0^{x_2}}\right)
-M_{B}^2 Q^2 t_0^2 {\cal
H}_{+12_4,-16_{2},-18_{4},-20_{4},+22_{16},-24_{20},-6_4}(x_i)+8 Q^2
t_0^3{\cal H}_{22}(x_i) \right.\right.\right. \nonumber\\&&
\left.\left.\left. +M_{B}^2 Q^2 \beta t_0^2 {\cal
H}_{-12_4,+16_{2},+18_{4},+20_{44},+22_{8},-24_{16},+6_4}(x_i)+Q^4\beta
t_0^4 {\cal H}_{+16_{4},+22_{16},-24_{32}}(x_i)\right.\right.\right.
\nonumber\\&& \left.\left. \left. \left.+Q^2\beta t_0^2 {\cal
H}_{-16_2,-22_{16},+24_{16}}(x_i)+M_{B}^2 Q^2 t_0^3 {\cal
H}_{-12_{20},+16_{2},+18_{2},-20_{46},-22_{32},+24_{20},-6_{12}}(x_i)
 \right.\right.\right.
\right.\nonumber\\&& \left.\left.\left.\left.  +M_{B}^2 Q^2
t_0^4{\cal
H}_{-14_{2},+15_{10},-16_{4},+18_{10},-2_{16},+20_{42},+21_{40},
+22_{8}, +23_{72},-24_{40},+4_{8},+6_{16},+8_{6},-9_{14}}(x_i)+
Q^2\beta t_0^3 {\cal H}_{22_{4}}(x_i) \right.\right.\right.\right.
\nonumber\\&& \left.\left.\left.\left.
 +M_{B}^2
Q^2\beta t_0^3 {\cal
H}_{+12_{20},-16_{2},-18_{14},-20_{170},-22_{16},+24_{16},+6_{12}}(x_i)
+Q^4t_0^4{\cal H}_{-16_{4},+22_{32},-24_{40}}(x_i)
\right.\right.\right.\right. \nonumber\\&&
\left.\left.\left.\left.+M_{B}^2 Q^2 \beta t_0^4 {\cal
H}_{-12_{16},+14_{2},-15_{10},+16_{4},+18_{14},-2_{8},+20_{122},+21_{32},
+22_{4},+23_{112},-24_{32},+4_{16},-6_{16},-8_{6},+9_{14}}(x_i)
\right.\right.\right.\right. \nonumber\\&& \left.\left.\left.
\left.+M_{B}^2 Q^2 t_0^5 {\cal
H}_{+12_{2},-16_{6},-18_{3},-20_{3},+22_{80},-24_{60},-6_{2}}(x_i)+Q^4
t_0^5{\cal H}_{+16_{4},-22_{48},+24_{40}}(x_i)+Q^4 \beta t_0^5 {\cal
H}_{-16_{4},-22_{24},+24_{32}}(x_i) \right.\right.\right.\right.
\nonumber\\&& \left.\left.\left.\left. +M_{B}^2 Q^2 \beta t_0^5
{\cal
H}_{-12_{2},+16_{6},+18,+20_{19},+22_{40},-24_{48},+6_{2}}(x_i)
+M_{B}^2 Q^2  t_0^6 {\cal
H}_{-12_{6},+16_{6},+18,-20_{11},-22_{72},+24_{60},-6_{2}}(x_i)
\right.\right.\right.\right. \nonumber\\&& \left.\left.\left.\left.
+M_{B}^2 Q^2 \beta t_0^6 {\cal
H}_{+12_{6},-16_{6},-18_5,-20_{55},-22_{36},+24_{48},+6_{2}}(x_i)+s(s_0,
Q^2)\left\{\vphantom{\int_0^{x_2}} Q^2 t_0^4 {\cal
H}_{-16_{4},+22_{32},-24_{40}}(x_i) \right.
\right.\right.\right.\right.\nonumber\
\end{eqnarray}
\begin{eqnarray}
&&+\left.\left.\left.\left.\left.   Q^2 t_0^4 \beta {\cal
H}_{+16_{4},+22_{16},-24_{32}}(x_i)+M_{B}^2 t_0^5 {\cal
H}_{+12_{2},-16_{6},-18_{3},-20_{3},+22_{48},-24_{60},-6_{2}}(x_i)-8Q^2
t_0^5{\cal H}_{22}(x_i)(2+\beta) \right. \right.\right.\right.
\right.\nonumber\\&& \left.\left.\left.\left.\left. +M_{B}^2 t_0^5
\left(\vphantom{\int_0^{x_2}}\beta {\cal
H}_{-12_{2},+16_{6},+18,+20_{19},+22_{24},-24_{48},+6_{2}}(x_i)+ t_0
{\cal H}_{-12_{4},-18,-20_{13},-22_{24},-6_{4}}(x_i) \right.
\right.\right. \right.\right.\right.\nonumber\\&&
\left.\left.\left.\left.\left.\left. +t_0\beta {\cal
H}_{+12_{4},-18_3,-20_{33},-22_{12},+6_{4}}(x_i)
 \vphantom{\int_0^{x_2}}\right)\vphantom{\int_0^{x_2}}\right\}
+\left\{\vphantom{\int_0^{x_2}}Q^4(-1+t_0) (1+2t_0^3)+Q^2M_{B}^2 t_0
\left(-1+ t_0-2t_0^2-3t_0^3+3t_0^4 \right) \right.\right.
\right.\right.\right.\nonumber\\&& \left.\left.\left.\left.\left.
-t_0^3(2Q^2+3M_{B}^2 t_0)s(s_0, Q^2)
 \vphantom{\int_0^{x_2}}\right\}\left[\vphantom{\int_0^{x_2}}
(1-\beta)\left(\vphantom{\int_0^{x_2}}6{\cal
H}_{10}(x_i)(t_0-x_2)+2x_2{\cal
H}_{16}(x_i)\vphantom{\int_0^{x_2}}\right)-4x_2{\cal
H}_{24}(x_i)(5+4\beta)
 \vphantom{\int_0^{x_2}}\right]
 \right.\right. \right.\right.\nonumber\\&& \left.\left.\left.\left.
 +4{\cal
H}_{22}(x_i)(2+\beta)\left[\vphantom{\int_0^{x_2}}Q^2
\left(\vphantom{\int_0^{x_2}}M_{B}^2(-1+t_0)
(2t_0-2t_0^2-t_0^3+9t_0^4)+Q^2(-2+4t_0-t_0^2-4t_0^3+6t_0^4)
\vphantom{\int_0^{x_2}}\right)
 \right.\right.\right. \right.\right.\nonumber\\&& \left.\left.\left.\left.\left.
+t_0^3(-2+t_0)(2Q^2+3M_{B}^2 t_0)s(s_0, Q^2)
\vphantom{\int_0^{x_2}}\right]
 \vphantom{\int_0^{x_2}}\right\}+\frac{2 m_N^4 m_b}{M_{B}^2 t_0}
 (-1+\beta)\left\{\vphantom{\int_0^{x_2}}{\cal
H}_{22}(x_i)(t_0-x_2)\left(\vphantom{\int_0^{x_2}}
Q^4(-2+3t_0-t_0^2-2t_0^3+6t_0^4) \right.\right.\right.\right.\right.
\nonumber\\&& \left.\left.\left.\left.\left.+M_{B}^2 Q^2
(-2t_0+3t_0^2-t_0^3-7t_0^4+9t_0^5)
+t_0^3(-1+t_0)(2Q^2+3t_0M_{B}^2)s(s_0,
Q^2)\vphantom{\int_0^{x_2}}\right)+Q^2
t_0^2\left[\vphantom{\int_0^{x_2}}-Q^2t_0{\cal
H}_{+16,-24_2}(x_i)\right.\right.\right.\right. \right.\nonumber\\&&
\left. \left.\left.\left.\left. -t_0^2 M_{B}^2{\cal
H}_{+12_2,+14,+15,+15,-20_4,+21_4-24_2,-6_2,+8,+9}(x_i)+t_0{\cal
H}_{10}(x_i) (Q^2+t_0 M_{B}^2)-x_2 (Q^2+t_0 M_{B}^2){\cal
H}_{+10,-16,+24_2}(x_i)
\vphantom{\int_0^{x_2}}\right]\vphantom{\int_0^{x_2}}\right\}
\right.\right.\right.\nonumber\\
&&+ \frac{m_N^3 Q^2}{M_{B}^2} \left\{\vphantom{\int_0^{x_2}}-M_{B}^2
Q^2(-1+\beta){\cal H}_{+12_2,-18_2,-6_2}(x_i)-M_{B}^2 Q^2 {\cal
H}_{20}(x_i)(2-22\beta)+M_{B}^2 Q^2 t_0(-1+\beta){\cal
H}_{+12_{10},-18,+6_6}(x_i) \right.  \nonumber\\&& \left. -M_{B}^2
Q^2 t_0{\cal H}_{20}(x_i)(23+85\beta)+Q^4 t_0^2 \beta {\cal
H}_{+10_{6},+16_2,+22_{8},-24_{16}}(x_i)+Q^4 t_0^2{\cal
H}_{-10_{6},-16_2,+22_{16},-24_{20}}(x_i)
 \right.  \nonumber\\&& \left.
+M_{B}^2 Q^2 t_0^2{\cal
H}_{-10_{6},+12-8,-14,+15_5,-16_2,+18_5,-2_8,+20_{21},+21_{20},+23_{36},
-24_{20},+4_{4},+6_{8},+8_{3},-9_{7}}(x_i) \right.  \nonumber\\&&
\left. +M_{B}^2 Q^2 t_0^2 \beta {\cal
H}_{+10_{6},-12_8,+14,-15_{5},+16_2,+18_7,-4_2,+20_{61},+21_{16},
+23_{56}, -24_{16},+4_{8},-6_{8},-8_{3},+9_{7}}(x_i)\right.
\nonumber\\&& \left. +M_{B}^2 Q^2 t_0^3{\cal
H}_{-10_{6},+12_4,-16_{2},-18_{6},-20_{6},+22_{32},-24_{20},-6_{4}}(x_i)+Q^4
t_0^3{\cal H}_{+10_{6},+16_2,-22_{24},+24_{20}}(x_i) \right.
\nonumber\\&& \left. +M_{B}^2 Q^2 t_0^3\beta {\cal
H}_{+10_{6},+12_4,+16_{2},+18_{2},+20_{38},+22_{16},-24_{16},
+6_{4}}(x_i)
 + Q^4 t_0^3\beta{\cal
H}_{-10_{6},-16_{2},-22_{12},+24_{16}}(x_i) \right.  \nonumber\\&&
\left. +M_{B}^2 Q^2 t_0^4{\cal
H}_{+10_{6},-12_{12},+16_{2},+18_{2},-20_{22},-22_{24},+24_{20},
-6_{4}}(x_i) +M_{B}^2 Q^2 t_0^4\beta{\cal
H}_{-10_{6},+12_{12},-16_{2},-18_{10},-20_{110},-22_{12},+24_{16},+6_{4}}(x_i)
\right.  \nonumber\\&& \left. +s(s_0,
Q^2)\left[\vphantom{\int_0^{x_2}} Q^2 t_0^2 {\cal
H}_{-10_{6},-16_{2},+22_{16},-24_{20}}(x_i)+Q^2 t_0^2 \beta {\cal
H}_{+10_{6},+16_{2},+22_{8},-24_{16}}(x_i)-4Q^2 t_0^3  {\cal
 H}_{22}(x_i)(2+\beta)
\right.\right.  \nonumber\\&& \left.\left. +M_{B}^2 t_0^3
  {\cal H}_{-10_{6},+12_{4},-16_{2},-18_{6},-20_{6},+22_{16},
 -24_{20},-6_{4}}(x_i) +M_{B}^2 t_0^3 \beta {\cal
 H}_{+10_6,-12_4,+16_2,+18_2,+20_{38},+22_8,-24_{16},+6_4}(x_i)
\right.\right.  \nonumber\\&& \left.\left.
 +M_{B}^2 t_0^4 {\cal
 H}_{-12_8,-18_2,-20_{26},-22_8,+6_{8}}(x_i)+M_{B}^2 t_0^4 \beta {\cal
 H}_{+12_8,-18_6,-20_{66},-22_4,+6_{8}}(x_i)
 \vphantom{\int_0^{x_2}}\right]+2t_0x_2\left[\vphantom{\int_0^{x_2}}2Q^2{\cal
 H}_{22}(x_i)(2+\beta)
\right.\right.  \nonumber\\&& \left.\left.
 \left(\vphantom{\int_0^{x_2}}M_{B}^2 t_0
 (-4+3t_0)+Q^2(-2+3t_0)\vphantom{\int_0^{x_2}}\right)-2Q^2{\cal
 H}_{24}(x_i)(5+4\beta) \left(\vphantom{\int_0^{x_2}}Q^2(-1+t_0)
 +M_{B}^2(-1-t_0+t_0^2)\vphantom{\int_0^{x_2}}\right)
\right.\right.  \nonumber\\&& \left.\left.
 +2\left(\vphantom{\int_0^{x_2}}{\cal
 H}_{24}(x_i)(5+4\beta)+{\cal
 H}_{22}(x_i)(2+\beta)(-2+t_0)\vphantom{\int_0^{x_2}}\right)
 (Q^2+M_{B}^2 t_0)s(s_0, Q^2)+{\cal
 H}_{+10_3,16}(x_i)(-1+\beta)
\right.\right.  \nonumber\\&& \left.\left.
\left(\vphantom{\int_0^{x_2}}Q^4(-1+t_0)
 +Q^2M_{B}^2(-1-t_0+t_0^2)-
 (Q^2+M_{B}^2 t_0)s(s_0, Q^2) \vphantom{\int_0^{x_2}}\right)
 \vphantom{\int_0^{x_2}}\right]
\vphantom{\int_0^{x_2}} \right\} -\frac{m_N^2 m_b
Q^2}{M_{B}^2}(-1+\beta)\left\{\vphantom{\int_0^{x_2}} Q^2 t_0
M_{B}^2\right. \nonumber\\&& \left.{\cal
 H}_{+12_2,+14,+15,-20_4,+21_4,-6_2,+8,+9}(x_i)+2 Q^2 t_0{\cal
 H}_{22}(x_i)\left(\vphantom{\int_0^{x_2}}Q^2(2-3t_0)+
 6M_{B}^2t_0(1-t_0)\vphantom{\int_0^{x_2}}\right)  -2{\cal
 H}_{22}(x_i)\left[\vphantom{\int_0^{x_2}}(-1+t_0)t_0
\right.\right.\nonumber\\&&\left.\left.
 (Q^2+M_{B}^2 t_0)s(s_0, Q^2)-x_2
 Q^4(-1+3t_0)+
 x_2(-1+t_0)\left(\vphantom{\int_0^{x_2}}M_{B}^2t_0s(s_0, Q^2)
 +Q^2(3M_{B}^2 t_0+s(s_0, Q^2))
  \vphantom{\int_0^{x_2}}\right)
 \vphantom{\int_0^{x_2}}\right]
\vphantom{\int_0^{x_2}}\right\} \nonumber\\&& +m_N Q^4
t_0\left\{\vphantom{\int_0^{x_2}}
Q^2\left(\vphantom{\int_0^{x_2}}2{\cal
 H}_{6}(x_i)(-1+\beta)(1+t_0)+{\cal
 H}_{18}(x_i)(-3+\beta+t_0-5t_0\beta)-{\cal
 H}_{20}(x_i)(3-19\beta+11t_0+55t_0\beta)
 \vphantom{\int_0^{x_2}}\right)
\right.\nonumber
\end{eqnarray}
\begin{eqnarray}&&
\left.
 -\left(\vphantom{\int_0^{x_2}}-2{\cal
 H}_{6}(x_i)(-1+\beta)(1+2t_0)+{\cal
 H}_{20}(x_i)(3-19\beta+13t_0+33t_0\beta)+{\cal
 H}_{18}(x_i)(3+t_0-\beta+3t_0\beta)\vphantom{\int_0^{x_2}}\right)
 s(s_0, Q^2)
\right.\nonumber\\&&\left.
 +2{\cal
 H}_{12}(x_i)(-1+\beta)\left(\vphantom{\int_0^{x_2}}
 Q^2(-1+3t_0)+(-1+2t_0)s(s_0, Q^2)
 \vphantom{\int_0^{x_2}}\right)
\vphantom{\int_0^{x_2}}\right\}
 \left.\vphantom{\int_0^{x_2}}\right\}
+\frac{m_N^2}{(Q^2+m_N^2t_0^2)^2 2\sqrt{6}}(t_0-x_2)
\left\{\vphantom{\int_0^{x_2}} -m_N
t_0\left[\vphantom{\int_0^{x_2}}{\cal
 H}_{16}(x_i)
\right.\right.\nonumber\\&&\left.\left.
 (-1+\beta)\left(\vphantom{\int_0^{x_2}}m_N^2(-1+t_0)+m_N m_b t_0
 + Q^2t_0(-1+t_0)-t_0s(s_0, Q^2)\vphantom{\int_0^{x_2}}\right)
 +{\cal
 H}_{10}(x_i)(-1+\beta)\left(\vphantom{\int_0^{x_2}}3m_N^2(-1+t_0)
 -m_N m_b t_0
\right.\right.\right.\nonumber\\&&\left.\left.\left.
 +3t_0 (Q^2(-1+t_0)-s(s_0, Q^2))\vphantom{\int_0^{x_2}}\right)
 -2{\cal
 H}_{24}(x_i)\left(\vphantom{\int_0^{x_2}}m_N^2(5+4\beta)(-1+t_0)
 +m_N^2 m_b(-1+\beta)t_0+(5+4\beta)t_0\left(\vphantom{\int_0^{x_2}}
 Q^2(-1+t_0)
\right.\right.\right.\right.\nonumber\\&&\left.\left. \left.\left.
 -s(s_0, Q^2)\vphantom{\int_0^{x_2}}\right)\vphantom{\int_0^{x_2}}
 \right)\vphantom{\int_0^{x_2}}
 \right]+{\cal
 H}_{22}(x_i)\left[\vphantom{\int_0^{x_2}}-m_N^2 m_b (-1+\beta)
 (-2+t_0)(-1+t_0)+2m_N^3(2+\beta)
 (2t_0-4t_0^2+t_0^3)-2m_N(2+\beta)
\right.\right.\nonumber\\&&\left.\left.
 t_0^2\left(\vphantom{\int_0^{x_2}}Q^2(-2+3t_0)+(-2+t_0)s(s_0, Q^2)
 \vphantom{\int_0^{x_2}}\right)+m_b(-1+\beta)
 t_0\left(\vphantom{\int_0^{x_2}}Q^2(-1+3t_0)+(-1+t_0)s(s_0, Q^2)\vphantom{\int_0^{x_2}}\right)
 \vphantom{\int_0^{x_2}}\right]
 \vphantom{\int_0^{x_2}}\right\}
\nonumber\\&& +\frac{m_N}{(Q^2+m_N^2t_0^2)
4\sqrt{6}M_{B}^2t_0}\left\{\vphantom{\int_0^{x_2}} 2m_N{\cal
 H}_{22}(x_i)(t_0-x_2)\left[\vphantom{\int_0^{x_2}}-m_N^2 m_b
 (-1+\beta)(-2+t_0)(-1+t_0)+2m_N^3(2+\beta)t_0
\right.\right.\nonumber\\&&\left.\left.
 (2-4t_0+t_0^2)-m_b
 (-1+\beta)t_0\left(\vphantom{\int_0^{x_2}}Q^2+2t_0M_{B}^2-3Q^2t_0
 +s(s_0, Q^2)+t_0s(s_0,
 Q^2)\vphantom{\int_0^{x_2}}\right)+2m_N(2+\beta)t_0^2
 \left(\vphantom{\int_0^{x_2}}Q^2(2-3t_0)
\right.\right.\right.\nonumber\\&&\left.\left.\left.
 +2t_0M_{B}^2+2s(s_0, Q^2)-t_0s(s_0, Q^2)
 \vphantom{\int_0^{x_2}}\right)
 \vphantom{\int_0^{x_2}}\right]
 +t_0\left[\vphantom{\int_0^{x_2}}-2m_N^2M_{B}^2{\cal
 H}_{18,+20,+6}(x_i)+m_N^2M_{B}^2\beta{\cal
 H}_{18_2,+20_{22},+6_2}(x_i)
\right.\right.\nonumber\\&&\left.\left.
 -m_N^4 t_0 {\cal
 H}_{+10_6,+16_{2},+20_{24}}(x_i)+m_N^2M_{B}^2 t_0{\cal
 H}_{+18,-20_{23},-6_{6}}(x_i)+m_N m_b M_{B}^2 t_0{\cal
 H}_{+14,+15,-20_{4},+21_{4},-6_{2},+8,+9}(x_i)
\right.\right.\nonumber\\&&\left.\left.
 -M_{B}^2 Q^2 t_0{\cal
 H}_{+18_3,-20_{3},-6_{2}}(x_i)+m_N^4 \beta t_0{\cal
 H}_{+10_6,+16_{2},-24_{16}}(x_i)-m_N^2M_{B}^2 t_0\beta {\cal
 H}_{-18_7,-20_{85},+6_{6}}(x_i)
\right.\right.\nonumber\\&&\left.\left.
 -m_N m_b M_{B}^2 t_0\beta{\cal
 H}_{+14,+15,-20_{4},+21_4,-6_2,+8,+9}(x_i)
 +M_{B}^2 Q^2 t_0\beta{\cal
 H}_{+18,+20_{19},+6_{2}}(x_i)+m_N^2 t_0^2(m_N^2-Q^2)
\right.\right.\nonumber\\&&\left.\left.
 {\cal
 H}_{+10_6,+16_{2},+24_{20}}(x_i)-m_N^3 m_b t_0^2{\cal
 H}_{+10_2,-16_{2},+24_{4}}(x_i)+m_N^2\beta t_0^2(Q^2-m_N^2){\cal
 H}_{+10_6,+16_{2},-24_{16}}(x_i)
\right.\right.\nonumber\\&&\left.\left.
 -m_N^2M_{B}^2 t_0^2{\cal
 H}_{+10_6,+14,-15_{5},+16_2,-18_{5},+8_{2},-20_{21},+21_{20},
 -23_{36},
 +24_{20},-4_{4},-6_{8},-8_{3},+9_{7}}(x_i)+m_N^3m_b\beta t_0^2{\cal
 H}_{+10_2,-16_{2},+24_{4}}(x_i)
\right.\right.\nonumber\\&&\left.\left.
 +M_{B}^2 Q^2 t_0^2{\cal
 H}_{+18,-20_{11},-6_{2}}(x_i)+ m_N^2M_{B}^2 t_0^2\beta{\cal
 H}_{+10_6,+14,-15_{5},+16_{2},+18_7,-2_4,+20_{61},+21_{16},+23_{56},
 -24_{16},+4_{8},+6_{8},-8_{6},+9_{7}}(x_i)
\right.\right.\nonumber\\&&\left.\left.
 -M_{B}^2 Q^2 \beta t_0^2{\cal
 H}_{+18_5,+20_{55},-6_{2}}(x_i)+m_N^2 Q^2 t_0^3{\cal
 H}_{+10_6,+16_2,+24_{20}}(x_i)-m_N^2 Q^2 \beta t_0^3{\cal
 H}_{+10_6,+16_2,-24_{16}}(x_i)
\right.\right.\nonumber\\
&&+s(s_0, Q^2)\left[\vphantom{\int_0^{x_2}}-M_{B}^2 t_0{\cal
 H}_{+18_3,+20_3,+6_{2}}(x_i)+M_{B}^2 \beta t_0{\cal
 H}_{+18,+20_{19},+6_{2}}(x_i)-m_N^2 t_0^2{\cal
 H}_{+10_6,+16_{2},+24_{20}}(x_i)
\right.\nonumber\\&&\left. -M_{B}^2 t_0^2{\cal
 H}_{+18,+20_{13},+6_{4}}(x_i) +m_N^2 t_0^2\beta{\cal
 H}_{+10_6,+16_{2},-24_{16}}(x_i)-M_{B}^2 t_0^2 \beta {\cal
 H}_{+18_3,+20_{33},-6_{4}}(x_i)\vphantom{\int_0^{x_2}}\right]
\nonumber\\&&
 -2M_{B}^2{\cal
 H}_{12}(x_i)(-1+\beta)\left(\vphantom{\int_0^{x_2}}m_N m_b t_0+
 m_N^2(-1+t_0)(-1+4t_0)+t_0Q^2-3Q^2t_0^2+
 t_0s(s_0, Q^2)-2t_0^2s(s_0, Q^2)\vphantom{\int_0^{x_2}}\right)
\nonumber\\&&+2x_2m_N^2{\cal
 H}_{16}(x_i)(-1+\beta)\left(\vphantom{\int_0^{x_2}}
 m_N^2(-1+t_0)+m_N m_b t_0 -t_0M_{B}^2-t_0Q^2+Q^2t_0^2
 -t_0s(s_0, Q^2)\left.\vphantom{\int_0^{x_2}}\right)
 \right.\nonumber\\&&\left.
+2x_2m_N^2{\cal
 H}_{10}(x_i)(-1+\beta)\left(\vphantom{\int_0^{x_2}}
 3m_N^2(-1+t_0)-m_N m_b t_0-3 t_0 (M_{B}^2+Q^2(1-t_0))
 -3t_0s(s_0, Q^2)
\vphantom{\int_0^{x_2}}\right) +4x_2m_N^2{\cal
 H}_{24}(x_i) \right.\nonumber\\&&\left.
\left(\vphantom{\int_0^{x_2}}-m_N^2(5+4\beta)(-1+t_0)
 -m_N m_b(-1+\beta)t_0+(5+4\beta)t_0
 \left(\vphantom{\int_0^{x_2}}M_{B}^2+Q^2(1-t_0)
 +s(s_0, Q^2)\vphantom{\int_0^{x_2}}\right)
 \vphantom{\int_0^{x_2}}\right)
 \left.\vphantom{\int_0^{x_2}}\right]
 \left.\vphantom{\int_0^{x_2}}\right\}
 \left.\vphantom{\int_0^{x_2}}\right]
 \left.\vphantom{\int_0^{x_2}}\right\}\right.\nonumber\\
&&\nonumber~~~~~~~~~~~~~~~~~~~~~~~~~~~~~~~~~~~~~~~~~~~~~~~~~~~~~~~~~~~~~~~~~~~~~~~~~~~~~~~~~~~~~~~~~~~~~~~~~~~~~~~~~~~~~~~~~~~~~~~~~~~~~~~~~~~~~~~~~(A.1)
\end{eqnarray}
 and
 \begin{eqnarray}\label{f_{2}}
&&f_{2}(Q^{2})= \frac{1}{2\sqrt{\lambda_{\Lambda_b}}}
e^{m_{\Lambda_b}^{2}/M_{B}^{2}}\left\{\vphantom{\int_0^{x_2}}
\int_{t_{0}}^{1} dx_{2}\int_{0}^{1-x_{2}}dx_{1}
e^{-s(x_{2},Q^{2})/M_{B}^{2}}\frac{1}{2\sqrt{6}x_2}
\left[\vphantom{\int_0^{x_2}}(-1 +\beta) {\cal
H}_{+11,+5}(x_i)+{\cal
H}_{+17}(x_i)(-1+6\beta)\vphantom{\int_0^{x_2}}\right]
\right.\nonumber\\&&\left.+\int_{t_0}^1dx_2\int_0^{1-x_2}dx_1
\int_{t_0}^{x_2}dt_1e^{-s(t_1,Q^2)/M_{B}^{2}
}\left\{\vphantom{\int_0^{x_2}} \frac{m_N^2}{M_{B}^{4}t1^2
2\sqrt{6}}\left[\vphantom{\int_0^{x_2}}-m_N m_b (-1 +\beta)x_2{\cal
 H}_{+10,-16,+24_{2}}(x_i)+2{\cal
 H}_{22}(x_i)
\right.\right.\right.\nonumber\\&&\left.\left.\left.
 \left(\vphantom{\int_0^{x_2}}m_N m_b(-1
+\beta)x_2 +(Q^2+s(t_1,Q^2))(-1 +\beta)x_2+m_N^2(1-\beta+\beta
x_2+5x_2)
 \vphantom{\int_0^{x_2}}\right)
\vphantom{\int_0^{x_2}}\right]+ \frac{m_N^2}{M_{B}^{4}t1 2\sqrt{6}}
\left[\vphantom{\int_0^{x_2}} m_N m_b (-1 +\beta)
\right.\right.\right.\nonumber\\&&\left.\left.\left. {\cal
 H}_{+10,-16,+24_{2}}(x_i)-2{\cal
 H}_{22}(x_i)\left(\vphantom{\int_0^{x_2}}m_N m_b (-1 +\beta)+
 (Q^2+s(t_1,Q^2))(-1 +\beta)-\left(\vphantom{\int_0^{x_2}}3Q^2+
 s(t_1,Q^2)(2 +\beta)
  \vphantom{\int_0^{x_2}}\right)x_2
\right.\right.\right.\right.\nonumber\\&&\left.\left.\left.\left.
  +m_N^2(5+\beta+2x_2+\beta x_2)
      \vphantom{\int_0^{x_2}}\right)
 \vphantom{\int_0^{x_2}}\right]
+ \frac{m_N^2}{M_{B}^{2}t1 4\sqrt{6}} \left[\vphantom{\int_0^{x_2}}
{\cal
 H}_{-14,+15_5,+18_{6},-8_2,+20_{10},+21_{20},-22_4,+36_{23},
 +4_4,+6_4,+8_3,-9_7}(x_i)
\right.\right.\right.\nonumber\\&&\left.\left.\left.
 -4{\cal
 H}_{12}(x_i)(-1 +\beta)+\beta {\cal
 H}_{+14,-15_5,+18_{6},-2_4,+20_{34},+21_{16},-22_8,+4_{8},
 -6_4,-8_3,+9_7}(x_i)-8x_2 (2+\beta){\cal
 H}_{22}(x_i)
\vphantom{\int_0^{x_2}}\right]
\right.\right.\nonumber\\&&\left.\left. + \frac{m_N^4}{M_{B}^{4}t1^3
\sqrt{6}}x_2(-1 +\beta){\cal
 H}_{22}(x_i)+ \frac{m_N}{M_{B}^{2}t1^2
2\sqrt{6}}\left(\vphantom{\int_0^{x_2}} m_b (-1 +\beta){\cal
 H}_{-12, +18,+20,6}(x_i)+2m_N x_2 (1 +2\beta){\cal
 H}_{22}(x_i)
\vphantom{\int_0^{x_2}}\right)
\right.\right.\nonumber\\&&\left.\left.
+\frac{m_N^2}{M_{B}^{4}\sqrt{6}}{\cal
 H}_{22}(x_i)\left(\vphantom{\int_0^{x_2}}
-3Q^2+m_N^2(2 +\beta)-s(t_1,Q^2)(2 +\beta)
 \vphantom{\int_0^{x_2}}\right)+\frac{2 m_N^2}{M_{B}^{2}\sqrt{6}}
 (2 +\beta){\cal H}_{22}(x_i)
\vphantom{\int_0^{x_2}}\right\} \right.\nonumber\\&&\left.
+\int_{t_{0}}^{1}dx_{2}\int_{0}^{1-x_{2}}dx_{1}e^{-s_{0}/M_{B}^{2}}
\left\{\vphantom{\int_0^{x_2}} (t_0-x_2)
\left[\vphantom{\int_0^{x_2}}
\frac{m_N^4t_0^2}{(Q^2+m_N^2t_0^2)^3\sqrt{6}}+
\frac{m_N^2}{(Q^2+m_N^2t_0^2)^2 2\sqrt{6}}
\vphantom{\int_0^{x_2}}\right]
 \left[\vphantom{\int_0^{x_2}} m_N m_b
t_0(-1 +\beta) \right.\right.\right.\nonumber\\&&\left.\left.\left.
{\cal
 H}_{+10, -16,+24,2}(x_i)+2{\cal
 H}_{22}(x_i) \left(\vphantom{\int_0^{x_2}}
-m_N m_b t_0 (-1 +\beta)-Q^2 t_0
(-1+\beta+3t_0)+m_N^2\left(\vphantom{\int_0^{x_2}}
1-\beta-t_0(5+\beta)+(2+\beta)t_0^2 \vphantom{\int_0^{x_2}}\right)
\right.\right.\right.\right.\nonumber\\&&\left.\left.\left.\left.
-t_0\left(\vphantom{\int_0^{x_2}}-1+\beta+(2+\beta)t_0
\vphantom{\int_0^{x_2}}\right)s(s_0,Q^2)
\vphantom{\int_0^{x_2}}\right)
 \vphantom{\int_0^{x_2}}\right]
 +\frac{m_N}{(Q^2+m_N^2t_0^2)M_{B}^{2}t_04\sqrt{6}}
\left[\vphantom{\int_0^{x_2}}4m_N {\cal
 H}_{22}(x_i)(t_0-x_2)\left(\vphantom{\int_0^{x_2}}-m_N m_b
 (-1 +\beta)t_0
\right.\right.\right.\right.\nonumber\\&&\left.\left.\left.\left.
 +m_N^2(1-\beta-5t_0-\beta t_0+2t_0^2+\beta
 t_0^2)+t_0\left(\vphantom{\int_0^{x_2}}
 Q^2-Q^2(\beta+3t_0)+M_{B}^{2}(-1-2\beta+2\beta t_0+4t_0)
 +s(s_0,Q^2)
\right.\right.\right.\right.\right.\nonumber\\&&\left.\left.\left.\left.\left.
 -(\beta+2t_0+\beta t_0)s(s_0,Q^2) \vphantom{\int_0^{x_2}}\right)
\vphantom{\int_0^{x_2}}\right)+t_0\left(\vphantom{\int_0^{x_2}}m_b
M_{B}^{2}{\cal
 H}_{-20_2, -6_2,+20_2,+6_2}(x_i)+m_N^2 m_b t_0 {\cal
 H}_{-10_2, +16_2,-24_4}(x_i)
\right.\right.\right.\right.\nonumber\\&&\left.\left.\left.\left.
 +M_{B}^{2} m_b t_0{\cal
 H}_{-14, +15_5,-2_8,+20_{10},+21_{20},+23_{36},+4_{4},+6_{4},
 +8_{3},-9_{7}}(x_i)+m_N^2 m_b \beta t_0 {\cal
 H}_{+10_2, -16_2,+24_4}(x_i)
\right.\right.\right.\right.\nonumber\\&&\left.\left.\left.\left.
 +M_{B}^{2}m_N \beta t_0{\cal
 H}_{+14, -15_5,-2_4,+20_{34},+21_{16},+23_{56},+4_{8},-6_{4},
 -8_{3},+9_{7}}(x_i)-2M_{B}^{2}{\cal
 H}_{12}(x_i)(-1 +\beta)(m_b+2m_N t_0)
\right.\right.\right.\right.\nonumber\\&&\left.\left.\left.\left.
 + 2M_{B}^{2}{\cal
 H}_{18}(x_i)\left(\vphantom{\int_0^{x_2}}m_b(-1 +\beta)+3m_Nt_0
 (1 +\beta)\vphantom{\int_0^{x_2}}\right)-2m_N^2 m_b x_2(-1 +\beta)
 {\cal
 H}_{+10,-16,+24_4}(x_i)
\vphantom{\int_0^{x_2}}\right)
 \vphantom{\int_0^{x_2}}\right]
 \vphantom{\int_0^{x_2}}\right\}
\vphantom{\int_0^{x_2}}\right\}
,\nonumber~~~~~~~~~~~~(A.2)
\end{eqnarray}

where
\begin{eqnarray}
{\cal H}(x_i) &=& {\cal H}(x_1,x_2,1-x_1-x_2),
\nonumber \\
s(y,Q^2)&=&(1-y)m_{N}^2+\frac{(1-y)}{y}Q^2+\frac{m_b^2}{y},~~~~~~~~~~~~~~~~~~~~~~~~~~~~~~~~~~~~~~~~~~~~~~~~~~~~~~~~~~~~~~~~~~~~~~~~~~~~~~~~~~~(A.3)\nonumber\
\end{eqnarray}
and $t_0=t_{0}(s_{0},Q^2)$ is the solution of the equation
$s(t_{0},Q^2)=s_{0}$, and is given as
\begin{eqnarray}
t_{0}(s_{0},Q^2)=\frac{m_N^2-Q^2+\sqrt{4m_N^2(m_b^2+Q^2)+(m_N^2-Q^2-s_0)^2}+s_0}{2m_N^2}.~~~~~~~~~~~~~~~~~~~~~~~~~~~~~~~~~~~~~~~~~~~~~~~~~~(A.4)\nonumber\
\end{eqnarray}

In calculations,   the following short hand notations for
the functions ${\cal H}_{\pm i_a,\pm j_b, ...}=\pm a{\cal H}_{i}\pm
b{\cal H}_{j}...$ have been used, and  ${\cal H}_{i}$ are given in terms of the
DA's as follows:
\begin{eqnarray}
&&{\cal H}_{1}=S_{1}~~~~~~~~~~~~~~~~~~~~~~~~~~~~~~~~~~~~~~~~{\cal
H}_{2}=S_{1,-2}\nonumber\\&&{\cal
H}_{3}=P_{1}~~~~~~~~~~~~~~~~~~~~~~~~~~~~~~~~~~~~~~~~{\cal
H}_{4}=P_{1,-2}\nonumber\\&&{\cal
H}_{5}=V_{1}~~~~~~~~~~~~~~~~~~~~~~~~~~~~~~~~~~~~~~~~{\cal
H}_{6}=V_{1,-2,-3}\nonumber\\&&{\cal
H}_{7}=V_{3}~~~~~~~~~~~~~~~~~~~~~~~~~~~~~~~~~~~~~~~~{\cal
H}_{8}=-2V_{1,-5}+V_{3,4}\nonumber\\&&{\cal
H}_{9}=V_{4,-3}~~~~~~~~~~~~~~~~~~~~~~~~~~~~~~~~~~~~~{\cal
H}_{10}=-V_{1,-2,-3,-4,-5,6}\nonumber\\&&{\cal
H}_{11}=A_{1}~~~~~~~~~~~~~~~~~~~~~~~~~~~~~~~~~~~~~~~{\cal
H}_{12}=-A_{1,-2,3}\nonumber\\&&{\cal
H}_{13}=A_{3}~~~~~~~~~~~~~~~~~~~~~~~~~~~~~~~~~~~~~~~{\cal
H}_{14}=-2A_{1,-5}-A_{3,4}\nonumber\\&&{\cal
H}_{15}=A_{3,-4}~~~~~~~~~~~~~~~~~~~~~~~~~~~~~~~~~~~~{\cal
H}_{16}=A_{1,-2,3,4,-5,6}\nonumber\\&&{\cal
H}_{17}=T_{1}~~~~~~~~~~~~~~~~~~~~~~~~~~~~~~~~~~~~~~~~{\cal
H}_{18}=T_{1,2}-2T_{3}\nonumber\\&&{\cal
H}_{19}=T_{7}~~~~~~~~~~~~~~~~~~~~~~~~~~~~~~~~~~~~~~~~{\cal
H}_{20}=T_{1,-2}-2T_{7}\nonumber\\&&{\cal
H}_{21}=-T_{1,-5}+2T_{8}~~~~~~~~~~~~~~~~~~~~~~~~~~{\cal
H}_{22}=T_{2,-3,-4,5,7,8}\nonumber \\&&{\cal
H}_{23}=T_{7,-8}~~~~~~~~~~~~~~~~~~~~~~~~~~~~~~~~~~~~~{\cal
H}_{24}=-T_{1,-2,-5,6}+2T_{7,8},~~~~~~~~~~~~~~~~~~~~~~~~~~~~~~~~~~~~~~~~~~~~~~~~~(A.5)\nonumber
\
\end{eqnarray}
 where for each DA's,  $X_{\pm i,\pm j, ...}=\pm X_{i}\pm X_{j}...$ have  also been used.

\end{document}